\documentclass[reprint, nofootinbib, amsmath, amssymb, aps, prl, preprintnumbers]{revtex4-2}

\usepackage{graphicx}
\usepackage{dcolumn}
\usepackage{bm}

\usepackage{physics}
\usepackage{mathrsfs}
\usepackage{amsmath, amsthm, amssymb}
\usepackage{amsfonts}
\usepackage{bbm}
\usepackage[hypertexnames=false]{hyperref} 
\usepackage{mathtools}
\usepackage{eucal}
\usepackage{bbm}
\usepackage{color}

\newcommand{\Eq}[1]{Eq.~(\ref{#1})}
\newcommand{\Eqs}[1]{Eqs.~(\ref{#1})}
\newcommand{\Fig}[1]{Fig.~{\ref{#1}}}

\newcommand{\ie}{\textit{i.e.~}}

\newcommand{\eg}{\textit{e.g.~}}

\newcommand{\Mp}{M_{\mathrm{P}}}

\begin{document}

\preprint{RESCEU-24/23}

\title{Why does inflation look single field to us?}

\author{Koki Tokeshi}
    \email{tokeshi@resceu.s.u-tokyo.ac.jp}
    \affiliation{
      Department of Physics and Research Center for the Early Universe (RESCEU), 
      Graduate School of Science, The University of Tokyo, Bunkyo, Tokyo 113-0033, Japan
    }
\author{Vincent Vennin}
    \email{vincent.vennin@phys.ens.fr}
    \affiliation{
    Laboratoire de Physique de l'\'{E}cole Normale Sup\'{e}rieure,
    ENS, CNRS, Universit\'{e} PSL, Sorbonne Universit\'{e},
    Universit\'{e} Paris Cit\'{e}, 75005 Paris, France
}

\date{\today}

\begin{abstract}
Most high-energy constructions that realise a phase of cosmic inflation contain many degrees of freedom. Yet, cosmological observations are all consistent with single-field embeddings. We show how volume selection effects explain this apparent paradox. Because of quantum diffusion, different regions of space inflate by different amounts. In regions that inflate most, and eventually dominate the volume of the universe, a generic mechanism is unveiled that diverts the inflationary dynamics towards single-field attractors. The formalism of constrained stochastic inflation is developed to this end.
\end{abstract}

\maketitle

\noindent{\em Introduction}.---Cosmic inflation~\cite{Sato:1980yn, Guth:1980zm, Starobinsky:1980te, Linde:1981mu, Albrecht:1982wi, Linde:1983gd} is a phase of accelerated expansion that occurred in the early universe, during which vacuum quantum fluctuations were amplified by gravitational instability and gave rise to density fluctuations on large scales. 
These fluctuations constitute the seeds of all cosmological structures, and the validity of this scenario has been confirmed by a wealth of high-precision astrophysical measurements, ranging from the temperature and polarisation fluctuations of the Cosmic Microwave Background (CMB)~\cite{Planck:2018nkj}, to galaxy and large-scale-structure surveys~\cite{SDSS:2005xqv, BOSS:2014hwf, EUCLID:2011zbd}. 
This makes inflation the leading paradigm to describe the early universe.   

Most physical setups that have been proposed to embed inflation contain a large number of high-energy degrees of freedom, see \eg \cite{Linde:1993cn, Kawasaki:2000yn, Lyth:2002my, Kachru:2003sx, Allahverdi:2006we, Kallosh:2004yh, Martin:2013tda, Baumann:2014nda, Vennin:2015vfa, Moultaka:2016frs, Weymann-Despres:2023wly, Kaiser:2013sna, DeCross:2015uza}. 
This is because inflation is expected to occur at very high energies, ranging from GeV to $10^{15}$ GeV scales.
The extensions to the standard model of particle physics that have been proposed to describe physics at those scales (supersymmetry, supergravity, string theory, etc.) usually come with many additional fields. 
The presence of multiple fields during inflation is expected to leave specific imprints, such as non-Gaussianities or entropic perturbations. However, all observations performed so far have failed to detect such features and are consistent with single-field models of inflation~\cite{Planck:2018jri, Planck:2019kim}. 
Therefore, a crucial question that remains open for inflation is: why do we observe single-field phenomenology, while inflation is expected to be realised in multiple-field setups?

In this article, we show how this question can be naturally answered by volume-selection effects. 
Because of quantum fluctuations, different regions in space inflate by different amounts. 
In a multiple-field landscape, it is shown that the regions that inflate the most reach an effectively single-field behaviour when the scales observed in the CMB are being produced. 
Since such regions expand their physical volume by a larger amount, they eventually dominate the content of the universe, which explains how single-field phenomenology emerges from multi-field setups. 

This mechanism, which plays a central role in explaining why inflation looks single field to us, is unveiled using the formalism of stochastic inflation~\cite{Starobinsky:1986fx}. 
In this approach, as quantum fluctuations cross out the Hubble radius during inflation, they become part of the large-scale classical fields and randomly shift the background configurations. 
The dynamics of the fields thus become stochastic at large scales, which in practice is described by Langevin equations. 
The time required to terminate inflation is promoted to a random variable, and its statistics can be studied using first-passage-time techniques. 
It is related to the observed curvature perturbation in the stochastic-$\delta \mathcal{N}$ formalism~\cite{Enqvist:2008kt, Fujita:2013cna, Vennin:2015hra}. 

In order to focus on the realisations of the Langevin equations that inflate for the longest period of time, and which thus dominate the volume of the universe, a direct approach consists in simulating a large number of realisations numerically and keeping only those that inflate more than a certain threshold. 
However, this method becomes prohibitively expensive when the threshold increases since most realisations are discarded. 
Moreover, since it relies on numerical sampling, it is not propitious for analytical insight. 
This is why, in this work, methods from the theory of ``constrained stochastic processes''~\cite{10.1063/1.3586036, Majumdar_2015, delarue2016conditioned} are borrowed, in order to derive modified Langevin equations that only generate realisations of a fixed duration (or a duration larger than a given bound).
This allows us to study the statistics of the original stochastic process, \emph{conditioned} to its duration. 

We find that imposing a long duration for inflation leads to large realisations of the noises to be sampled at early time, which drifts the system towards the lightest field direction. 
This explains why, starting from initial conditions that would normally result in substantial multiple-field signatures to be produced, predictions are aligned with the single-field behaviour once the selection on the duration of inflation is applied. 
We also show that the early phase of large-noise realisations lies in general outside the observable window, hence it does not threaten the viability of the currently preferred models of inflation. 

Natural units are used with $c = \hbar = 1$ and $\Mp = 1 / \sqrt{8 \pi G} \simeq 2.4 \times 10^{18} \, \mathrm{GeV}$ denotes the reduced Planck mass.\\

$ $\\{\em Stochastic inflation}.---On a homogeneous and isotropic background, fields are conveniently expanded into Fourier modes, 
\begin{align}
    \vb*{\phi} (\vb*{x}, \, N) 
    &= \vb*{\phi}^{-} (\vb*{x}, \, N) + \vb*{\phi}^{+} (\vb*{x}, \, N) \,\, , 
    \label{eq:decomp}
    \\ 
    \vb*{\phi}^{\pm} (\vb*{x}, \, N) 
    &\equiv \int \frac{ \dd^3 \vb*{k} }{ (2 \pi)^3 } 
    \, \Theta \qty[ \pm \qty( k - \sigma a H) ] 
    \, {\vb*{\phi}}_{\vb*{k}}  (N) 
    \, e^{i \vb*{k} \cdot \vb*{x}} \,\, . 
    \notag 
\end{align}
Here, $\vb*{\phi}$ is a vector that contains all field configurations and momenta, and time is labeled with the number of $e$-folds $N$, related to cosmic time $t$ through $\dd N = H \, \dd t$, where $H = \dot{a} / a$ is the Hubble parameter and $a$ the scale factor. 
The coarse-grained field $\vb*{\phi}^{-}$ contains all wavelengths larger than the Hubble radius $1 / H$ (rescaled by the constant parameter $\sigma$), while $\vb*{\phi}^{+}$ contains smaller wavelengths. 
As comoving Fourier modes cross out the Hubble radius, they go from $\vb*{\phi}^{+}$ to $\vb*{\phi}^{-}$, and contribute a white Gaussian noise $\vb*{\xi}$ to the dynamics of $\vb*{\phi}^{-}$ (simply denoted $\vb*{\phi}$ hereafter) that reads~\cite{Starobinsky:1986fx} 
\begin{equation}
    \dv{\vb*{\phi}}{N} 
    = \vb*{F} (\vb*{\phi}) + \vb*{G} (\vb*{\phi}) \, \vb*{\xi} (N) \,\, . 
    \label{eq:si_lan}
\end{equation}
Here, $\vb*{F}$ describes the homogeneous (in the limit $\sigma \ll 1$), classical dynamics of the fields, while the amplitude of the noise $\vb*{G}$ is obtained from evolving quantised cosmological perturbations from the Bunch--Davies vacuum at small scales. 
For a scalar field $\phi$ in the slow-roll regime, $F = - V' / 3H^2$ and $G = H / 2\pi$, where $H^2 = V / 3\Mp^2$ and $V$ is the potential energy of the fields, while the field momentum is set by the slow-roll attractor~\cite{Grain:2017dqa}. Inflation terminates when $\ddot{a}$ stops being positive. 
This defines a final hypersurface $\mathcal{C}$ in the field space on which absorbing boundary conditions are imposed.  

The Langevin equation~(\ref{eq:si_lan}) gives rise to a Fokker--Planck equation for the probability density $P(\vb*{\phi}, \, N)$ of $\vb*{\phi}$ at time $N$,
\begin{equation}
    \pdv{P}{N} 
    = \qty( 
        - \grad \cdot \vb*{F} + \frac{1}{2} \grad \otimes \grad \vcentcolon \vb*{G}^2 
    ) P 
    \,\, . 
    \label{eq:fp}
\end{equation}
Hereafter, It\^o's convention is adopted for explicitness (although extensions to other discretisation conventions are straightforward~\cite{Pinol:2020cdp}) and Frobenius inner product's notation $\grad \otimes \grad \vcentcolon \vb*{G}^2 = (\partial^2 / \partial \phi^{i} \partial \phi^{j})  {G^{j}}_{k} {G^{ki}}$ is used where $\grad$ is the field space gradient. 

Starting from the initial condition $\vb*{\phi}$, the number of $e$-folds elapsed until the first crossing of $\mathcal{C}$ is a random variable denoted by  $\mathcal{N}$ and referred to as the first-passage time. 
Its distribution function, $P_{\mathrm{FPT}} (\vb*{\phi}, \, \mathcal{N})$, obeys the adjoint Fokker--Planck equation~\cite{Vennin:2015hra,Pattison:2017mbe}
\begin{equation}
    \pdv{P_{\mathrm{FPT}}} {\mathcal{N}} 
    = \qty(
        \vb*{F} \cdot \grad + \frac{1}{2} \vb*{G}^2 \vcentcolon \grad \otimes \grad 
    ) P_{\mathrm{FPT}} \equiv \mathcal{L} P_{\mathrm{FPT}} \,\, .
    \label{eq:fp_adj}
\end{equation}
This partial differential equation needs to be solved with the boundary condition $P_{\mathrm{FPT}} (\vb*{\phi}, \, \mathcal{N}) = \delta_{\mathrm{D}} (\mathcal{N})$ for $\vb*{\phi} \in \mathcal{C}$, where $\delta_{\mathrm{D}}$ is the Dirac distribution. 
At large $\mathcal{N}$, the upper tail of $P_{\mathrm{FPT}}$ decays exponentially with $\mathcal{N}$~\cite{Pattison:2017mbe, Ezquiaga:2019ftu}, which makes the sampling of long-lasted realisations numerically challenging.\\

$ $\\{\em Constrained stochastic processes}.---Consider the subset of realisations of the stochastic process~\eqref{eq:si_lan} starting from $\vb*{\phi}_{0}$ that realise a fixed number of $e$-folds $N_{\mathrm{F}}$. 
At time $N$, they follow a distribution function that is denoted by $\mathcal{P} (\vb*{\phi}, \, N \mid N_{\mathrm{F}})$. 
Using Bayes theorem, this can be written in terms of quantities defined above in the unconstrained process,
 \begin{equation}
    \mathcal{P} (\vb*{\phi}, \, N \mid N_{\mathrm{F}}) 
    = \frac{ 
        P_{\mathrm{FPT}} (\vb*{\phi}, \, N_{\mathrm{F}} - N) 
        P(\vb*{\phi}, \, N) 
    }{ 
        P_{\mathrm{FPT}} ( \vb*{\phi}_{0}, \, N_{\mathrm{F}} ) 
    } 
    \,\, . 
    \label{eq:csr_calp}
\end{equation}
Here, we used that, for Markovian processes, the probability to realise a total $N_{\mathrm{F}}$ $e$-folds if the system is at $\vb*{\phi}$ at time $N$, is equal to the probability to realise $N_{\mathrm{F}} - N$ $e$-folds starting from $\vb*{\phi}$. 
Since $P$ and $P_{\mathrm{FPT}}$ satisfy \eqref{eq:fp} and \eqref{eq:fp_adj} respectively, the above leads to~\cite{supp_ref} 
\begin{equation}
    \pdv{\mathcal{P}}{N} 
    =
     \qty( 
        - \grad \cdot \widetilde{\vb*{F}} 
        + \frac{1}{2} \grad \otimes \grad \vcentcolon \vb*{G}^2 
    ) \mathcal{P} 
    \,\, ,
    \label{eq:csr_fpeq}
\end{equation}
where
\begin{equation}
    \widetilde{\vb*{F}} (\vb*{\phi}, \, N) 
    \equiv \vb*{F} (\vb*{\phi}) + \vb*{G}^2 (\vb*{\phi}) \grad \ln P_{\mathrm{FPT}} (\vb*{\phi}, \, N_{\mathrm{F}} - N) \,\, . 
    \label{eq:csr_drift}
\end{equation}
One notices that \eqref{eq:csr_fpeq} is of the same form as \eqref{eq:fp}, \textit{i.e.} it is a Fokker--Planck equation, except that the drift function $\widetilde{\vb*{F}}$ now explicitly depends on time. 
As such, \eqref{eq:csr_fpeq} can equivalently be written in the Langevin form
\begin{equation}
    \dv{\vb*{\phi}}{N} 
    = \widetilde{\vb*{F}} (\vb*{\phi}, \, N) + \vb*{G} (\vb*{\phi}) \, \vb*{\xi} (N) \,\, ,
\label{eq:csr_lan}
\end{equation}
which is known as the Doob's transformation of (\ref{eq:si_lan})~\cite{Doob1957ConditionalBM}. 

The additional term in (\ref{eq:csr_drift}) is an effective force induced by the selection effect. 
When $N < N_{\mathrm{F}}$, the boundary condition, $P_{\mathrm{FPT}} = \delta_{\mathrm{D}}$ on $\mathcal{C}$, implies that the induced force is infinitely repelling on the final surface, preventing realisations from finishing before $N_{\mathrm{F}}$. 
As $N$ approaches $N_{\mathrm{F}}$, since $P_{\mathrm{FPT}} (\vb*{\phi}, \, \mathcal{N}) \propto \exp [ - f (\vb*{\phi}) / \mathcal{N} ]$ for small time arguments,
where $f$ grows with the distance between $\vb*{\phi}$ and $\mathcal{C}$~\cite{supp_ref}, the induced force becomes infinitely attracting towards the final surface and inflation necessarily terminates at $N_{\mathrm{F}}$. 
This guarantees that the duration of inflation is indeed fixed in the constrained process. 
Let us stress that sampling (\ref{eq:si_lan}) and keeping only realisations that produce $N_{\mathrm{F}}$ $e$-folds is mathematically equivalent to sampling (\ref{eq:csr_lan}).

A regime of physical interest below is when the noise amplitude $\vb*{G}$ is small. 
In that limit, $P_{\mathrm{FPT}} (\mathcal{N})$ is approximately Gaussian, 
\begin{equation}
    P_{\mathrm{FPT}} (\vb*{\phi}, \, \mathcal{N}) 
    \simeq \frac{1}{\sqrt{2 \pi \sigma^2 (\vb*{\phi})}} \, 
    \exp \qty{ 
        - \frac{ [ \mathcal{N} - \mu (\vb*{\phi}) ]^2 }{ 2 \sigma^2 (\vb*{\phi}) } 
    } 
    \,\, , 
    \label{eq:Pfpt:Gaussian:main}
\end{equation}
where $\mu$ is the mean number of $e$-folds and $\sigma$ is its standard deviation. 
From \eqref{eq:fp_adj}, they satisfy $\mathcal{L} \mu = -1$ and $\mathcal{L} \sigma^2 = - \vb*{G}^2 \vcentcolon (\grad \mu) \otimes (\grad \mu )$, with $\mathcal{L} \simeq \vb*{F} \cdot \grad$ at leading order in the noise. 
This gives rise to the induced drift
\begin{equation}
    \widetilde{\vb*{F}} 
    = \vb*{F} + \frac{ N_{\mathrm{F}} - N - \mu }{\sigma^2} \vb*{G}^2 \cdot
    \qty( 
        \grad \mu + \frac{N_{\mathrm{F}} - N - \mu}{2\sigma^2} \grad \sigma^2 
    ) 
    \,\, . 
\end{equation}
Here, $\sigma$ scales like $\vb*{G}$, hence the induced force is independent of the noise amplitude. 
In that regime, the noise term in \eqref{eq:csr_lan} can thus be neglected~\cite{supp_ref}, and the constrained dynamics becomes quasi-deterministic. 
Note that, from the point of view of the unconstrained dynamics, the noise plays a crucial role in the realisations that produce $N_{\mathrm{F}}$ $e$-folds, especially if $N_{\mathrm{F}}$ differs substantially from the mean first-passage time. 
However, in the regime described here, statistical fluctuations \emph{among the subset of constrained realisations} are negligible. 
This is why the noise can be neglected in \eqref{eq:csr_lan}, but obviously not in \eqref{eq:si_lan}. 

Stochastic processes of a fixed duration are dubbed ``excursions''~\cite{Majumdar_2015}, but other constrained processes can be sampled similarly, \eg ``meanders'' in which only a lower bound on the duration of inflation, $\mathcal{N} \geq N_{\mathrm{F}}$, is imposed. 
In that case, one still obtains a modified Langevin equation of the form~(\ref{eq:csr_lan}), except that $P_{\mathrm{FPT}} (\vb*{\phi}, \, N_{\mathrm{F}} - N)$ in (\ref{eq:csr_drift}) needs to be replaced with $\displaystyle \int_{N_{\mathrm{F}} - N}^{\infty} \dd \mathcal{N} \, P_{\mathrm{FPT}} (\vb*{\phi}, \, \mathcal{N})$~\cite{supp_ref}. 
In the low-diffusion limit, whether the volume selection effect during inflation is implemented via excursions or meanders leads to similar conclusions~\cite{supp_ref}, hence in the following we focus on excursions.\\

$ $\\{\em Single-field phenomenology from multi-field inflation}.---Let us now apply the formalism presented above to multiple-field models of inflation. 
For concreteness, a double quadratic potential is considered~\cite{Polarski:1992dq, Langlois:1999dw, Rigopoulos:2005us, Rigopoulos:2005xx}, 
\begin{equation}
    V (\vb*{\phi}) 
    = \frac{m_{1}^{2}}{2} \phi_{1}^{2} + \frac{m_{2}^{2}}{2} \phi_{2}^{2} 
    \,\, , 
    \label{eq:df_pot}
\end{equation}
where $\sqrt{r} \equiv m_2 / m_1>1$ is assumed without loss of generality.  
Upon introducing the rescaled variables, $x \equiv \phi_{1} / \Mp$, $y \equiv \phi_{2} / \Mp$, and $v_{0} \equiv m_{1}^{2} / 24 \pi^2 \Mp^2$, in the slow-roll regime the unconstrained Langevin equation~\eqref{eq:si_lan} reads
\begin{equation}
    \dv{N} \mqty( x \\  y ) 
    = - \frac{2}{x^2 + r y^2} \mqty( x \\ r y ) + \sqrt{v_0 (x^2 + r y^2)} \, \vb*{\xi} (N) 
    \,\, .
    \label{eq:df_lan}
\end{equation}
Inflation terminates on the contour $\mathcal{C}$ defined by $2 (x^2 + r^2y^2) = (x^2 + ry^2)^2$, on which an absorbing boundary is placed. 
When the stochastic noise is neglected in \eqref{eq:df_lan}, inflation proceeds along the classical lines $y = c x^r$, where $c$ is conserved and depends on the initial condition. 
This leads to the classical number of $e$-folds, $\mu (x, \, y) = (x^2 + y^2 - x_{\mathrm{F}}^2 - y_{\mathrm{F}}^2) / 4$, where $(x_{\mathrm{F}}, \, y_{\mathrm{F}})$ is the intersection of the classical trajectory with $\mathcal{C}$, and thus implicitly depends on the initial conditions $x$ and $y$.

Following \cite{PhysRevD.83.023522}, the prevalence of multi-field effects can be assessed by comparing the rate at which the trajectory turns in field space, $\eta_{\perp}$, with the rate at which it accelerates, $\eta_{\parallel}$. 
In other words, if $\vb*{v} = (\dd x / \dd N, \, \dd y / \dd N)^{\mathrm{T}}$ denotes the field-space velocity, $\eta_{\parallel}$ is the component of $\dd \vb*{v} / \dd N$ that is parallel to $\vb*{v}$, and $\eta_{\perp}$ is its orthogonal component. 
In the model~\eqref{eq:df_pot}, one finds 
\begin{equation}
\label{eq:lambda:w}
    \lambda 
    \equiv \frac{\eta_{\perp}}{\eta_{\parallel}} 
    = r (r-1) \frac{ w (1 + r w^2) }{ r^4 w^4 - (r^3 - 4 r^2 + r) w^2 + 1 } 
    \,\, , 
\end{equation}
where $w \equiv y/x$. 
In what follows, quasi single-field phenomenology will be associated with regions where $\lambda < \lambda_{\mathrm{c}}$, with $\lambda_{\mathrm{c}}$ a fixed threshold on which our conclusions do not depend. 
When the two fields have the same mass, $\lambda = 0$ and the setup is effectively single-field. 
For a large mass ratio, $r \gg 1$, single-field phenomenology requires $w < \lambda_{\mathrm{c}} / r^2$, which delineates a smaller field-space region as $r$ increases. 

\begin{figure*}
    \centering
    \includegraphics[width = 0.99\linewidth]{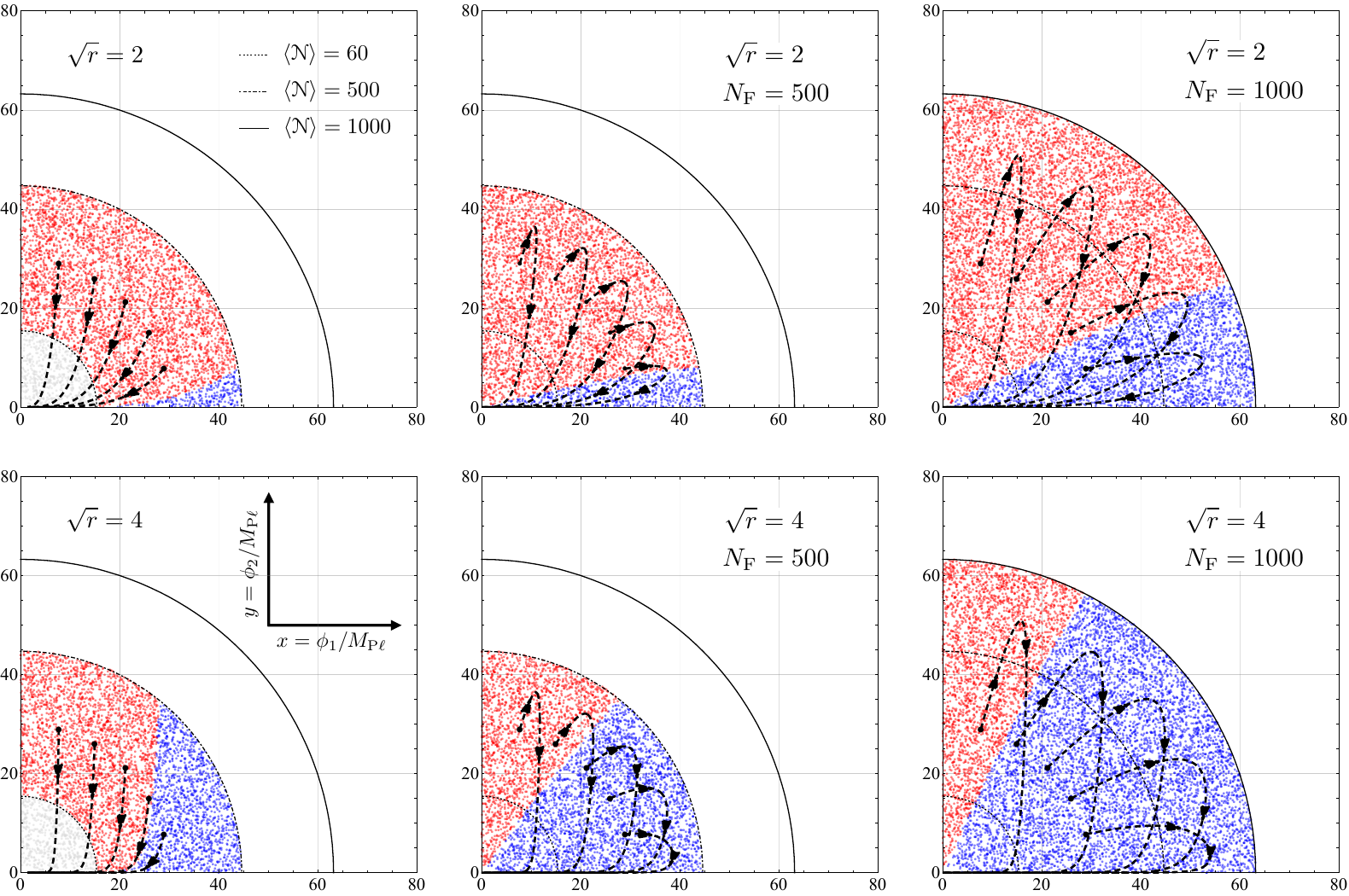}
    \caption{
        Random initial conditions are drawn inside a contour of equal $e$-folds, from which the inflationary dynamics is solved. 
        $\lambda$ is computed $60$ $e$-folds before the end of inflation, and the initial condition is marked in blue if $\lambda < \lambda_{\mathrm{c}} = 0.1$ (single-field phenomenology), and in red otherwise. 
        A few examples of inflationary trajectories are displayed with the black thick dashed lines, the initial conditions of which are set on the contour $x^{2} + y^{2} = 30^{2}$ (hence $\mu = \expval{ \mathcal{N} } = 225.5$ $e$-folds are classically realised). 
        The left panels correspond to the unconstrained setup, where the grey dots realise less than $60$ $e$-folds. 
        The middle and right panels show the constrained setup. 
        When $N_{\mathrm{F}}$ increases, more initial conditions give rise to single-field phenomenology (see also Fig. S5 in~\cite{supp_ref}). 
    }
    \label{fig:df_traj}
\end{figure*}
  
From a given initial condition, one can integrate the classical dynamics and compute $\lambda$ when the scales probed in the CMB emerge, \ie $60$ $e$-folds before the end of inflation. 
If $\lambda < \lambda_{\mathrm{c}}$, this initial condition is said to yield quasi single-field phenomenology, and is displayed in blue in the left panels of \Fig{fig:df_traj}. 
One can see that, as $r$ increases, the space of initial conditions compatible with single-field phenomenology is enlarged, although the single-field condition becomes more stringent as mentioned above. 
This is because, when the heavy field is heavier, it gets more quickly suppressed during inflation, and the system is more efficiently attracted towards the light field direction. 
Nonetheless, a fair fraction of the initial conditions yield multi-field effects. 

Let us now turn on the selection effect, and impose that $N_{\mathrm{F}}$ $e$-folds are realised. 
In practice, CMB measurements~\cite{Planck:2018jri} constrain $v_0$ to be of order $10^{-13}$, hence the noise amplitude in~\eqref{eq:df_lan} is highly suppressed and the Gaussian approximation \eqref{eq:Pfpt:Gaussian:main} can be used. 
There, $\mu$ was given above, and one finds $\sigma^2 \simeq v_0 [f(x,y) - f(x_{\mathrm{F}}, \, y_{\mathrm{F}})] / 48$, where $f (x, \, y) = x^6 + r y^6 + 3 [r (r+2) / (2r + 1)] x^2 y^4 + 3 [ (2r+1) / (r+2) ] x^4 y^2$. 
At leading order in $v_0$, the constrained dynamics is deterministic, so one can apply the same procedure as in the classical unconstrained setup and the result is displayed in the middle and right panels of \Fig{fig:df_traj}. 
One can see that, if one imposes $\mathcal{N} = N_{\mathrm{F}}$, the realisations of the noise that are selected divert the system towards a large detour at larger-field values. 
This detour circles clockwise, such that the system is much closer to the light-field direction when it approaches the end of inflation than what it would be without selection effects. 
As a consequence, one notices that more initial conditions give rise to single-field phenomenology once selection effects are turned on. 

When $N_{\mathrm{F}}$ increases, the detour is wider and the preference for single-field phenomenology is more pronounced. 
Although this effect is shown explicitly in the two-field model~\eqref{eq:df_pot}, we expect it to be generic. Indeed, in the unconstrained setup, the stochastic noise may either take the system closer to the heavy-field direction or to the light-field direction. 
If the system gets closer to the heavy field, the subsequent number of $e$-folds it realises is smaller than if it gets closer to the light field. 
As a consequence, imposing a large duration of inflation necessarily results in a biased sampling in favour of those noise realisations that bring the system closer to the light field, \ie to the field-space regions with stronger single-field phenomenology. 
This generic mechanism is present in any multiple-field model.\\

$ $\\{\em When do selection effects take place?}---Volume selection diverts the inflationary dynamics towards single-field looking regions, and a natural question is whether or not this detour leaves observable effects. 
Indeed, the predictions of inflationary models are usually derived along the unconstrained trajectory, and one may wonder how they change when computed along the constrained ones. 
This question already arises in single-field models, so for simplicity $\phi_2$ is set to $0$ in \eqref{eq:df_pot}. 

\begin{figure}
    \centering
    \includegraphics[width = 0.95\linewidth]{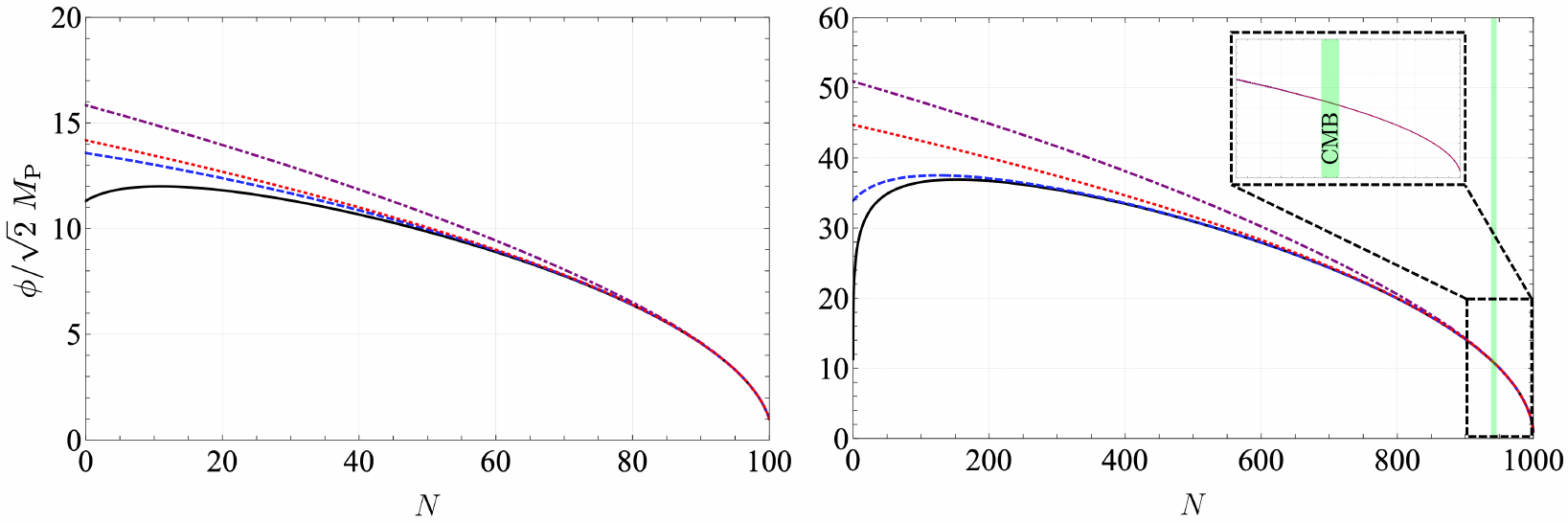}
    \caption{
        Constrained inflationary trajectories, which become deterministic in the regime $v_0 \ll 1$, from several initial conditions when imposing $N_{\mathrm{F}} = 100$ (\textit{left}) and $1000$ (\textit{right}). 
        The red dotted line corresponds to the unconstrained trajectory that leads to $N_{\mathrm{F}}$ $e$-folds of inflation. 
    }
  \label{fig:sf_traj}
\end{figure}

In that case, the above formulas can be used with setting $y = 0$. 
The first-passage-time problem can be solved semi-analytically~\cite{supp_ref}, but when $v_0\ll 1$, the Gaussian approximation~\eqref{eq:Pfpt:Gaussian:main} applies and the constrained dynamics becomes deterministic. 
It is displayed in \Fig{fig:sf_traj}, where one can see that by $60$ $e$-folds before the end of inflation, the constrained realisations have all collapsed to the unconstrained one. 
This means that the part of the inflationary era that is probed in cosmological surveys is not affected by selection effects, and confirms that standard results apply. 

The reason why selection effects mostly take place early on is because quantum diffusion is more prominent at early stages of inflation (in the present model, $\vb*{F}$ decays as $1/x$ while $\vb*{G}$ grows as $x$). 
Since it is more likely for the system to fluctuate away from the mean path when the amplitude of the noise is larger, this explains why the selection detour is imprinted at early stages, \ie at scales larger than those observed~\cite{footnote_conf}.
In the vast majority of single-field models~\cite{Martin:2013tda} quantum diffusion is larger at earlier times, but there exist potentials that feature a transient phase of large diffusion at late time, and which are relevant to primordial black hole production for instance. 
The presence of observable imprints from selection effects in these setups would be interesting to further explore.\\

$ $\\{\em Conclusion}.---We have developed the formalism of constrained random processes in the context of stochastic inflation. 
This allowed us to derive effective Langevin equations that select the realisations where quantum diffusion leads to the longest duration of inflation. 
The corresponding space-time regions dominate the volume of the universe at late time and thus represent the most likely past history of a given observer.

We have found that, in multiple-field setups, these regions are efficiently diverted to single-field attractors. 
As a consequence, the set of initial conditions that leads to single-field phenomenology is much larger than in the absence of volume-selection effects. 
This mechanism is generic and helps to explain why, although inflation is most commonly realised in high-energy constructions that involve multiple additional degrees of freedom, cosmological observations are compatible with single-field models. 
If multi-field signatures are detected in the future, it would point towards the limited class of models where, 60 e-folds before the end of inflation, multi-field effects are produced everywhere in field space.

We have also shown that selection effects take place at early time and barely affect the last $60$ $e$-folds of inflation. 
This implies that the standard predictions of inflation are unaffected once the single-field attractor has been reached (although possible observable imprints of selection effects in models with substantial quantum diffusion towards the end of inflation~\cite{Ando:2020fjm, Tada:2021zzj}, or in the case of ``just-enough inflation''~\cite{Ramirez:2011kk, Cicoli:2014bja}, require further investigations). 

Note that, contrary to previous attempts to implement volume-weighting procedures~\cite{Sasaki:1988df, Nambu:1988je, Nakao:1988yi, Nambu:1989uf, Linde:1993xx}, our results do not depend on a choice of a volume measure. 
Instead, it relies on selecting the realisations that inflate the most, which are shown to reach single-field attractors at late time. 
Single-field phenomenology would therefore emerge for any volume measure, only the strength of the attractor would depend on the details of that measure. 

Finally, let us mention that the formalism of constrained random processes can be used more generally to sample rare realisations, in complement with other importance-sampling methods~\cite{Jackson:2022unc}. This may be of practical interest \eg in situations leading to the formation of primordial black holes. 

$ $\\
{\em Acknowledgment}.---The authors are grateful to  David Dean and Yuichiro Tada for fruitful discussions. K.~T.~thanks LPENS for hospitality and JSPS for support under KAKENHI Grant No.~21J20818. 

\bibliography{apssamp}

\clearpage
\onecolumngrid 
\begin{center}
   \textbf{\large Supplemental Material for ``Why does inflation look single field to us?''}\\[.2cm]
   Koki Tokeshi${}^{1, *}$ and Vincent Vennin${}^{2, \dagger}$\\[.2cm]
   {
      \itshape 
      ${}^1$Department of Physics and Research Center for the Early Universe (RESCEU),\\
      Graduate School of Science, The University of Tokyo, Bunkyo, Tokyo 113-0033, Japan
      \\[.1cm]
      ${}^2$Laboratoire de Physique de l'\'{E}cole Normale Sup\'{e}rieure,
      ENS, CNRS, \\
      Universit\'{e} PSL, Sorbonne Universit\'{e},
      Universit\'{e} Paris Cit\'{e}, 75005 Paris, France\\[.2cm]
   }
   Electronic address: 
   ${}^{*}$\href{mailto:tokeshi@resceu.s.u-tokyo.ac.jp}{tokeshi@resceu.s.u-tokyo.ac.jp}, 
   ${}^{\dagger}$\href{mailto:vincent.vennin@phys.ens.fr}{vincent.vennin@phys.ens.fr}
   \\[.2cm]
   (Dated: \today)\\[1cm]
\end{center}

\makeatletter
\renewcommand{\thesection}{S\arabic{section}}
\renewcommand{\theequation}{\thesection.\arabic{equation}}
\renewcommand{\thefigure}{S\arabic{figure}}
\renewcommand{\thetable}{S\arabic{table}}
\makeatother
\setcounter{section}{1}
\setcounter{equation}{0}
\setcounter{figure}{0}
\setcounter{table}{0}
\setcounter{page}{1}
\makeatletter

\section{S1.~Derivation of Constrained Equations}

A stochastic process described by the Langevin equation 
\begin{equation}
    \dv{\vb*{\phi}}{N} 
    = \vb*{{F}} (\vb*{\phi}) + \vb*{G} (\vb*{\phi}) \, \vb*{\xi} (N) 
    \,\, , 
    \label{eq:si_lan:SM}
\end{equation}
 is considered, terminating on the field-phase space hypersurface $\mathcal{C}$. 
Here, $\vb*{\xi} (N) $ is a white Gaussian noise with $\expval{ \vb*{\xi} (N) } = \vb*{0}$ and $\expval{ \vb*{\xi} (N_{1}) \vb*{\xi}^{\mathrm{T}} (N_{2}) } = \delta_{\mathrm{D}} (N_{1} - N_{2}) \mathbbm{1}$. 
Using It\^o's convention, this can be recast as a Fokker--Planck equation~\cite{risken1996fokker} for the probability $P(\vb*{\phi}, \, N)$ to find the system with the configuration $\vb*{\phi}$ at time $N$, 
\begin{equation}
    \pdv{P (\vb*{\phi}, \, N)}{N} 
    = \qty( 
        - \grad \cdot \vb*{F} + \frac{1}{2} \grad \otimes \grad \vcentcolon \vb*{G}^2 
    ) 
    P (\vb*{\phi}, \, N)\equiv \mathcal{L}^\dagger P(\vb*{\phi}, \, N)
    \,\, . 
    \label{eq:fp:SM}
\end{equation}
This defines the field-phase space differential operator $\mathcal{L}$, which appears through its adjoint in order to match the convention usually employed in the literature. The adjoint is defined in the usual sense, \ie for any field-phase space functions $f(\vb*{\phi})$ and $g(\vb*{\phi})$
\begin{equation}
\displaystyle \int \dd \vb*{\phi} \, \left[\mathcal{L} f(\vb*{\phi})\right] g(\vb*{\phi}) = \int \dd \vb*{\phi} \, f(\vb*{\phi}) \left[\mathcal{L}^\dagger g(\vb*{\phi})\right]\, . 
\end{equation}
In \Eq{eq:fp:SM}, Frobenius inner product's notation $\grad \otimes \grad \vcentcolon \vb*{G}^2 = (\partial^2 / \partial \phi^{i} \partial \phi^{j})  {G^{j}}_{k} G^{ki}$ is used, where $\grad$ is the field-phase space gradient and we use Einstein summation convention. 
Our goal is to derive an evolution equation for the probability associated with the system's configuration, under a certain condition imposed \eg on the total duration of the realisation, and starting from the initial configuration $\vb*{\phi}_{0}$. 
This probability density is denoted by $\mathcal{P} (\vb*{\phi}, \, N \mid C)$, where $C$ is a condition to be specified.  
Three types of condition will be considered, following the terminology of \cite{Majumdar_2015} and generalising their results.

\begin{center}
-----------------
\end{center}

\noindent\textbf{\textit{Excursions}}~~
When the overall duration of the realisation is fixed, the constrained process is called an ``excursion''. 
In this case, let $P [ \vb*{\phi}, \, N \cap \mathcal{N} (\vb*{\phi}_{0})  = N_{\mathrm{F}} ]$ denote the joint probability that, starting from $\vb*{\phi}_{0}$, the system is at $\vb*{\phi}$ at time $N$, \emph{and} the total duration of the realisation is $N_{{\mathrm{F}}}$. 
By using the rules of conditional probabilities, this can be written as the probability to realise $N_{{\mathrm{F}}}$ $e$-folds, multiplied by the probability that the system is at $\vb*{\phi}$ at time $N$ under the condition that $\mathcal{N} (\vb*{\phi}_{0}) = N_{{\mathrm{F}}}$. 
In other words,
\begin{equation}
    P [ \vb*{\phi}, \, N \cap \mathcal{N} (\vb*{\phi}_{0}) = N_{{\mathrm{F}}} ] 
    = \mathcal{P} [ \vb*{\phi}, \, N \mid \mathcal{N} (\vb*{\phi}_{0}) = N_{{\mathrm{F}}} ] 
    P_{{\mathrm{FPT}}} (\vb*{\phi}_{0}, \, N_{\mathrm{F}}) 
    \,\, ,
\end{equation}
where $P_{{\mathrm{FPT}} } (\vb*{\phi}, \, \mathcal{N})$ denotes the first-passage-time distribution function for the number of $e$-folds $\mathcal{N}$ realised from $\vb*{\phi}$. 
The ordering of the conditions can be swapped, \ie one can write $P [ \vb*{\phi}, \, N \cap \mathcal{N} (\vb*{\phi}_{0}) = N_{{\mathrm{F}}}]$ as the probability for the system to be at $\vb*{\phi}$ at time $N$, multiplied by the probability to realise a total of $N_{{\mathrm{F}}}$ $e$-folds under the condition that the system crosses $\vb*{\phi}$ at time $N$. 
Since the process is Markovian, the latter is nothing but the probability to realise $N_{{\mathrm{F}}} - N$ $e$-folds starting from $\vb*{\phi}$, hence
\begin{equation}
    P [ \vb*{\phi}, \, N \cap \mathcal{N} (\vb*{\phi}_0) = N_{{\mathrm{F}}} ] 
    = P_{{\mathrm{FPT}} } ( \vb*{\phi}, \, N_{{\mathrm{F}}} - N ) 
    P (\vb*{\phi}, \, N)
    \,\, .
\end{equation}
By equating the two equations above, one obtains
\begin{equation}
    \mathcal{P} [ \vb*{\phi}, \, N \mid \mathcal{N} (\vb*{\phi}_{0}) = N_{{\mathrm{F}}} ] 
    = \frac{ 
        P_{{\mathrm{FPT}} } (\vb*{\phi}, \, N_{{\mathrm{F}} } - N ) P (\vb*{\phi}, \, N)
    }{
        P_{{\mathrm{FPT}} } (\vb*{\phi}_{0}, \, N_{\mathrm{F}})
    }
    \,\, .
    \label{eq:constrainedP:excursion}
\end{equation}

\noindent\textbf{\textit{Meanders}}~~
If only a lower bound is imposed on the duration of inflation, $\mathcal{N} (\vb*{\phi}_{0}) \geq N_{{\mathrm{F}}}$, the constrained process is called a ``meander''. 
In that case, following similar lines as for the excursion, one can write
\begin{equation}
    P [ \vb*{\phi}, \, N \cap \mathcal{N} (\vb*{\phi}_{0}) \geq N_{{\mathrm{F}}} ] 
    = \mathcal{P} [ \vb*{\phi}, \, N \mid \mathcal{N} (\vb*{\phi}_{0}) \geq N_{{\mathrm{F}}} ] P [ \mathcal{N} (\vb*{\phi}_{0}) \geq N_{{\mathrm{F}}} ] 
    \,\, ,
    \label{eq:meandre:interm1}
\end{equation}
where 
\begin{equation}
    P [ \mathcal{N} (\vb*{\phi}_{0}) \geq N_{{\mathrm{F}}} ] 
    = \int_{N_{{\mathrm{F}}}}^{\infty} \dd \mathcal{N} \, P_{{\mathrm{FPT}}} (\vb*{\phi}_{0}, \, \mathcal{N}) 
    \label{eq:survival:def}
\end{equation}
is the probability that the duration of inflation, starting from $\vb*{\phi}_{0}$, exceeds $N_{{\mathrm{F}}}$. 
Using again that the stochastic process is Markovian, one can also write
\begin{equation}
    P [\vb*{\phi}, \, N \cap \mathcal{N} (\vb*{\phi}_{0}) \geq N_{{\mathrm{F}}} ] 
    = P [ \mathcal{N} (\vb*{\phi}) \geq N_{{\mathrm{F}}} - N ] 
    P (\vb*{\phi}, \, N) 
    \,\, .
    \label{eq:meandre:interm2}
\end{equation}
By equating \Eqs{eq:meandre:interm1} and~\eqref{eq:meandre:interm2}, one obtains
\begin{equation}
    \mathcal{P} [ \vb*{\phi}, \, N \mid \mathcal{N} (\vb*{\phi}_{0}) \geq N_{{\mathrm{F}}} ] 
    = \frac{
        P [ \mathcal{N} (\vb*{\phi}) \geq N_{{\mathrm{F}}} - N ] P (\vb*{\phi}, \, N)
    }{
        P [\mathcal{N} (\vb*{\phi}_{0}) \geq N_{{\mathrm{F}}} ] 
    } 
    \,\, .
    \label{eq:constrainedP:meander}
\end{equation} 

\noindent\textbf{\textit{Bridges}}~~
Finally, although it is not of direct relevance for inflationary dynamics, let us mention the case of stochastic bridges, where the condition is that $\vb*{\phi}(N_{{\mathrm{F}}}) \in \mathcal{C}$, but [contrary to excursions] the system is allowed to cross $\mathcal{C}$ at earlier times too. 
In that case, one has
\begin{equation}
    P [ \vb*{\phi}, \, N \cap \vb*{\phi} (N_{{\mathrm{F}}}) \in \mathcal{C} ] 
    = \mathcal{P} [ \vb*{\phi}, \, N \mid \vb*{\phi}(N_{{\mathrm{F}}}) \in \mathcal{C} ] P [ \vb*{\phi} (N_{{\mathrm{F}}}) \in \mathcal{C} \mid \vb*{\phi}_{0} ] 
    \,\, ,
    \label{eq:bridge:interm1}
\end{equation}
where $P [ \vb*{\phi} (N) \in \mathcal{C} \mid \vb*{\phi} ]$ is the probability that, starting from $\vb*{\phi}_{0}$ at time $N = 0$, the system lies in $\mathcal{C}$ at time $N$, \ie 
\begin{equation}
    P [ \vb*{\phi} (N)\in \mathcal{C} \mid \vb*{\phi}_{0} ] 
    = \int_{\mathcal{C}} \dd \vb*{\phi} \, P (\vb*{\phi}, \, N) 
    \,\, , 
    \label{eq:backward:Kolmogorov}
\end{equation}
with $P (\vb*{\phi}, \, N)$ being the solution of \Eq{eq:fp:SM} with initial condition $P (\vb*{\phi}, \, N = 0) = \delta_{\mathrm{D}} (\vb*{\phi} - \vb*{\phi}_{0})$. 
Using the fact that the stochastic process is Markovian, one can also write
\begin{equation}
    P [ \vb*{\phi}, \, N \cap \vb*{\phi} (N_{{\mathrm{F}}}) \in \mathcal{C} ] 
    = P [ \vb*{\phi} (N_{{\mathrm{F}}} - N) \in \mathcal{C} \mid \vb*{\phi} ] 
    P (\vb*{\phi}, \, N) 
    \,\, , 
    \label{eq:bridge:interm2}
\end{equation}
and upon identifying \Eqs{eq:bridge:interm1} and~\eqref{eq:bridge:interm2} one finds
\begin{equation}
    \mathcal{P} [ \vb*{\phi}, \, N \mid \vb*{\phi} (N_{{\mathrm{F}}}) \in \mathcal{C} ] 
    = \frac{
        P [ \vb*{\phi} (N_{{\mathrm{F}}} - N) \in \mathcal{C} \mid \vb*{\phi} ] 
        P (\vb*{\phi}, \, N )
    }{
        P [ \vb*{\phi} (N_{{\mathrm{F}}}) \in \mathcal{C} \mid \vb*{\phi}_{0} ] 
    } 
    \,\, . 
    \label{eq:constrainedP:bridge}
\end{equation}
 
\begin{center}
-----------------
\end{center}
 
It is interesting to notice that, in all three cases, the constrained probability density of the field is of the form
\begin{equation}
    \mathcal{P} (\vb*{\phi}, \, N \mid C ) 
    = \frac{ 
        Q (\vb*{\phi}, \, N_{{\mathrm{F}}} - N ) P (\vb*{\phi}, \, N) 
    }{ 
        Q (\vb*{\phi}_{0}, \, N_{\mathrm{F}})
    } 
    \,\, , 
    \label{eq:constrainedP:gen}
\end{equation}
with a function $Q$ that depends on the constraint under consideration. 
For excursions, $Q (\vb*{\phi}, \, N) = P_{{\mathrm{FPT}}} (\vb*{\phi}, N)$ is nothing but the first-passage-time distribution $P_{{\mathrm{FPT}}}$. 
For meanders, $Q (\vb*{\phi}, \, N) = P [ \mathcal{N} (\vb*{\phi}) \geq N ]$ is the so-called ``survival'' probability defined in \Eq{eq:survival:def}. 
For bridges, $Q (\vb*{\phi}, N) = P [ \vb*{\phi} (N) \in \mathcal{C} \mid \vb*{\phi} ]$ is the backward Kolmogorov distribution defined in \Eq{eq:backward:Kolmogorov}. 
Moreover, in all three cases, the $Q$ function obeys the adjoint Fokker--Planck equation, 
\begin{equation}
    \pdv{Q (\vb*{\phi}, \, N)}{N} 
    = \qty(\vb*{F} \cdot \grad + \frac{1}{2} \vb*{G}^2 \vcentcolon \grad \otimes \grad 
    ) Q(\vb*{\phi}, \, N) 
    = \mathcal{L} Q(\vb*{\phi}, \, N)
    \,\, .
\end{equation}
In excursions, this follows from the fact that $P_{{\mathrm{FPT}}}$ obeys the adjoint Fokker--Planck equation, see \Eq{eq:fp_adj}. 
In meanders, this can again be seen as a direct consequence of \Eq{eq:fp_adj} when applying the adjoint Fokker--Planck operator to \Eq{eq:survival:def}. 
In bridges, this is because the backward Kolmogorov distribution obeys the adjoint Fokker--Planck equation as well~\cite{risken1996fokker}. 
Therefore, the $Q$ functions only differ by the boundary conditions they need to be solved with.

This makes the derivation of an evolution equation for $\mathcal{P} (\vb*{\phi}, \, N \mid C)$ possible, since using the fact that $P$ satisfies the Fokker--Planck equation and that $Q$ satisfies the adjoint Fokker--Planck equation, differentiating \Eq{eq:constrainedP:gen} with respect to time leads to
\begin{equation}
    \pdv{ \mathcal{P}(\vb*{\phi}, \, N)}{N} 
    = - \frac{ P ( \vb*{\phi}, \, N) }{ Q (\vb*{\phi}_{0}, \, N_{{\mathrm{F}}}) }   \mathcal{L} Q (\vb*{\phi}, \, N_{\mathrm{F}}-N)
    + \frac{ Q (\vb*{\phi}, \, N_{\mathrm{F}}-N) }{ Q (\vb*{\phi}_{0}, \, N_{\mathrm{F}}) } \mathcal{L}^\dagger P (\vb*{\phi}, \, N) 
    \,\, .
    \label{eq:constrained:interm}
\end{equation}
To simplify this expression, it is useful to operate $\mathcal{L}^{\dagger}$ on the product $QP$ [where the arguments of these field-phase space functions are omitted for convenience], and this gives 
\begin{align}
    \mathcal{L}^{\dagger} (QP) 
    &= \qty[ 
            - \pdv{\phi^i} F^{i} (\vb*{\phi}) + \frac{1}{2} \frac{\partial^2}{\partial \phi^{i} \partial \phi^{j}} {G^{j}}_{k} (\vb*{\phi})  G^{ki} (\vb*{\phi}) 
        ] Q ( \vb*{\phi} ) P  ( \vb*{\phi} ) \notag \\ 
    &= Q \mathcal{L}^{\dagger} P - P \mathcal{L} Q 
    + \frac{1}{2} \pdv{\phi^{j}} ( {G^{j}}_{k} G^{ki} P ) \pdv{\phi^{i}} Q
    + \qty( \pdv{\phi^{i}} \pdv{\phi^{j}} Q ) {G^{j}}_{k} G^{ki} P + \frac{1}{2} \pdv{\phi^{i}} ( {G^{j}}_{k} G^{ki} P ) \pdv{\phi^{j}} Q \notag \\ 
    &= Q \mathcal{L}^{\dagger} P - P \mathcal{L} Q + \pdv{\phi^{j}} ({G^{j}}_{k} G^{ki} P ) \pdv{\phi^{i}} Q
    + \qty( \pdv{\phi^{i}} \pdv{\phi^{j}} Q ) {G^{j}}_{k} G^{ki} P \notag \\ 
    &= Q \mathcal{L}^{\dagger} P - P \mathcal{L} Q + \pdv{\phi^{i}} \qty( {G^{j}}_{k} G^{ki} P \pdv{\phi^{j}} Q ) \notag \\  
    &= Q \mathcal{L}^{\dagger} P - P \mathcal{L} Q + \pdv{\phi^{i}} \Bigg[ 
        \underbrace{ \left( {G^{j}}_{k} G^{ki} \pdv{\ln Q}{\phi^{j}} \right)}_{\equiv \Delta F^{i}} QP 
    \Bigg] 
    \,\, ,  
    \label{eq:supmat_qp}
\end{align}
where the last expression defines the vector $\Delta \vb*{F} = \vb*{G}^2 \vb*{\nabla} \ln Q(\vb*{\phi}, \, N_{{\mathrm{F}}} - N)$. 
Inserting the above into \Eq{eq:constrained:interm}, one finds
\begin{align}
    \pdv{ \mathcal{P} (\vb*{\phi}, \, N)}{N} 
    &= \frac{1}{Q (\vb*{\phi}_{0}, \, N_{{\mathrm{F}}})} 
    [
        Q (\vb*{\phi}, \, N_{{\mathrm{F}}} - N ) \mathcal{L}^{\dagger} P (\vb*{\phi}, \, N) 
        - P (\vb*{\phi}, \, N) \mathcal{L} Q (\vb*{\phi}, \, N_{{\mathrm{F}}} - N) 
    ] \notag\\
    &= \frac{1}{ Q ( \vb*{\phi}_{0}, \, N_{{\mathrm{F}}} )} \{ 
        \mathcal{L}^{\dagger} [ 
            Q (\vb*{\phi}, \, N_{{\mathrm{F}}} - N ) P (\vb*{\phi}, \, N) 
        ] - \grad \cdot [ 
            \Delta \vb*{F} Q (\vb*{\phi}, \, N_{{\mathrm{F}}} - N) P (\vb*{\phi}, \, N) 
        ]  
    \} \notag\\
    &= \mathcal{L}^{\dagger} [ 
        \mathcal{P} (\vb*{\phi}, \, N) 
    ] - \grad \cdot [ 
        \Delta \vb*{F} \mathcal{P} (\vb*{\phi}, \, N) 
    ] \notag\\
    &= \qty[ 
        - \grad \cdot (\vb*{F} + \Delta \vb*{F}) + \frac{1}{2} \grad \otimes \grad \vcentcolon \vb*{G}^2 
    ] \mathcal{P} (\vb*{\phi}, \, N) 
    \,\, .
\end{align}
The structure of this equation is identical to \Eq{eq:fp:SM}. 
As a consequence, the constrained process follows a Fokker--Planck equation where only the drift is modified and is given by
\begin{equation}
    \widetilde{\vb*{F}} 
    = \vb*{F} + \Delta \vb*{F} 
    = \vb*{F} + \vb*{G}^2 \cdot \grad \ln Q ( \vb*{\phi}, \, N_{{\mathrm{F}}} - N ) 
    \,\, .
\end{equation} 
Equivalently, it can be written in the Langevin form
\begin{equation}
    \dv{\vb*{\phi}}{N} 
    = \vb*{F} (\vb*{\phi}) + \Delta \vb*{F} (\vb*{\phi}, \, N) + \vb*{G} (\vb*{\phi}) \, \vb*{\xi} (N) 
    \,\, .
    \label{eq:si_lan:constrained:SM}
\end{equation}
\section{S2.~Example: Flat Potential}
\setcounter{section}{2}
\setcounter{equation}{0}
In order to illustrate the formalism introduced above, let us apply it to the case where inflation is driven by a single scalar field $\phi$ in a flat potential 
$V(\phi) = V_0 = \mathrm{const}$. 
It is assumed that inflation occurs at $\phi \geq \phi_{\mathrm{B}}$, and terminates when $\phi$ reaches $\phi_{\mathrm{B}}$. 
After rescaling the inflaton as 
\begin{equation}
    x \equiv 2 \pi \sqrt{ \frac{3\Mp^2}{V_{0}} } ( \phi - \phi_{\mathrm{B}} ) 
    \,\, ,
\end{equation} 
the Langevin equation reads
\begin{equation}
\label{eq:Lang:flat}
    \frac{\dd x}{\dd N} = \xi(N) 
    \,\, , 
\end{equation} 
with an absorbing boundary at $x = 0$. 
In the notations introduced above, $F = 0$ and $G = 1$. 
The Fokker--Planck equation is thus given by
\begin{equation}
    \pdv{ P (x, \, N) }{N} = \frac{1}{2} \pdv[2]{P (x, \, N)}{N} 
    \,\, , 
    \label{eq:FP:flat}
\end{equation} 
and can be solved by the method of images. 
Indeed, since it describes a free diffusion process, it admits Gaussian solutions, and the combination of Gaussian solutions that satisfies the initial condition $P(x, \, N = 0) = \delta_{\mathrm{D}} (x - x_0)$ together with the absorbing boundary condition $P(x = 0, \, N) = 0$ is given by
\begin{equation}
    P(x, \, N) 
    = \frac{1}{\sqrt{2 \pi N}} \, \qty{ 
        \exp \qty[ - \frac{ (x - x_{0})^2 }{2 N} ] - \exp \qty[ - \frac{ (x + x_{0})^2 }{2 N} ] 
    } \Theta (x) \,\, , 
    \label{eq:FP:sol:flat} 
\end{equation}
where $\Theta$ is the Heaviside function, \ie $\Theta (x) = 1$ if $x \geq 0$ and $0$ otherwise. 

From this expression, the survival probability $S(x_0, \, N)$, \ie the probability to be still inflating at time $N$, can be obtained as 
\begin{equation}
   \label{eq:flat:survival}
    S(x_0, \, N) 
    = \int_{0}^{\infty} \dd x \, P(x, \, N) 
    = \mathrm{erf} \qty( \frac{x_{0}}{\sqrt{2 N}} ) 
    \,\, .
\end{equation}
As explained around \Eq{eq:survival:def}, this is nothing but the probability for the first-passage time to be larger than $N$, so $\displaystyle S(x_{0}, \, N) = \int_{N}^{\infty} \dd \mathcal{N} \, P_{{\mathrm{FPT}}} (x_0, \, \mathcal{N})$. 
The first-passage-time distribution can thus be obtained as $P_{{\mathrm{FPT}}} (x_{0}, \, \mathcal{N}) 
= - \partial S / \partial N$, which leads to
\begin{equation}
    P_{{\mathrm{FPT}}} (x_{0}, \, \mathcal{N}) 
    = \frac{x_{0}}{\sqrt{2\pi} \, \mathcal{N}^{3/2}} \, \exp \qty( - \frac{x_{0}^2}{2 \mathcal{N}} ) 
    \,\, . 
    \label{eq:Levy}
\end{equation}
This is called a L\'{e}vy distribution. 
One can check that it satisfies the adjoint Fokker--Planck equation, which in the present case is the same as the Fokker--Planck equation~\eqref{eq:FP:flat}. 
\\ 
\begin{figure}
    \centering
    \includegraphics[width = 0.99\linewidth]{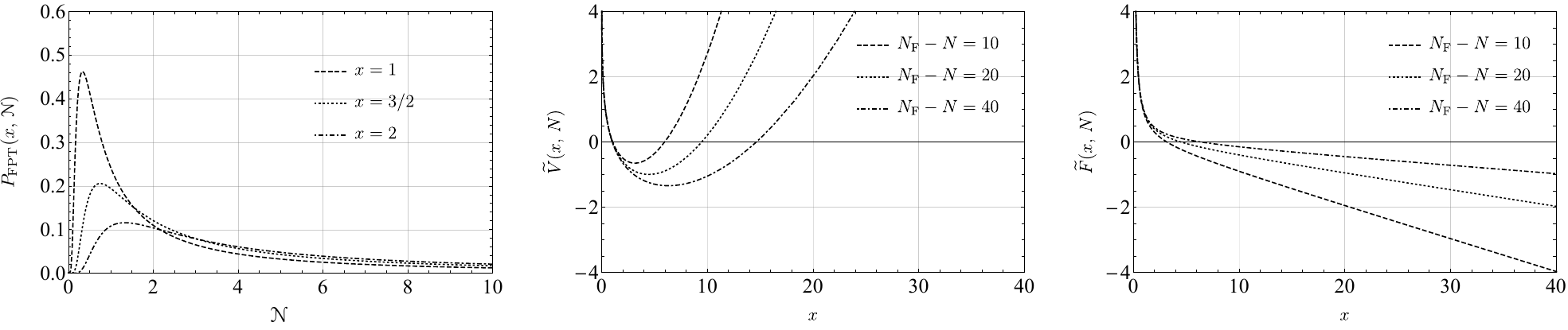}
    \caption{Flat-potential model~\eqref{eq:Lang:flat}. \textit{Left}: First-passage time distribution (\ref{eq:Levy}) for a few initial locations of $x$. \textit{Middle}: Effective potential (\ref{eq:effectivepot:excursion:flat}) in the excursion dynamics.  \textit{Right}: Induced force (\ref{eq:Finduced:excursion:bridge}) in the excursion dynamics. 
    }
    \label{fig:supmat_typ}
\end{figure}
\\
\noindent\textbf{\textit{Excursions}}~~
In the case of an excursion, the $Q$ function is nothing but the first-passage-time distribution, so 
\begin{equation}
      \widetilde{F} (x, \, N) 
      = \pdv{x} \ln \qty{ 
            \frac{x}{\sqrt{2\pi} (N_{{\mathrm{F}}} - N)^{3/2}} \, \exp \qty[ - \frac{x^2}{2 (N_{{\mathrm{F}}} - N)} ] 
        } 
      = \frac{1}{x} - \frac{x}{N_{{\mathrm{F}}} - N} 
      \,\, .
      \label{eq:Finduced:excursion:bridge}
\end{equation}
When $x$ approaches the absorbing boundary, the first term in this expression diverges, which prevents inflation from terminating before $N_{{\mathrm{F}}}$. 
In contrast, the second term pushes the field towards the absorbing boundary, and since it diverges when $N$ approaches $N_{{\mathrm{F}}}$, it localises all realisations on the absorbing boundary at the final time $N_{{\mathrm{F}}}$. 
This behaviour can also be seen at the level of the potential $\widetilde{V}$ from which the induced force $\widetilde{F} = - \partial \widetilde{V} / \partial  x$ derives,
\begin{equation}
    \widetilde{V} (x, \, N) 
    = - \ln x +  \frac{x^2}{2 ( N_{{\mathrm{F}}} - N )} 
    \,\, .
    \label{eq:effectivepot:excursion:flat}
\end{equation}
This potential is minimal at $x_{\mathrm{c}} = \sqrt{N_{{\mathrm{F}}} - N}$, which asymptotes to $0$ when $N \to N_{{\mathrm{F}}}$. 
Therefore, as $N$ proceeds, the field is attracted towards the absorbing boundary. 
The effective mass around that minimum is $\widetilde{V}'' (x_{\mathrm{c}}) = 2 / (N_{{\mathrm{F}}} - N)$, which diverges as $N$ approaches $N_{{\mathrm{F}}}$, impliying that the field becomes strongly confined around $x_{\mathrm{c}}$, hence around the absorbing boundary, see \Fig{fig:supmat_typ}. 

Let us note that the divergence of the attracting component of the induced force when $N \to N_{{\mathrm{F}}}$ is related to the fact that the first-passage-time distribution vanishes when $\mathcal{N} \to 0$, and that it does so exponentially fast, $P_{{\mathrm{FPT}}} (x, \, \mathcal{N}) \propto \exp [- f(x) / \mathcal{N}]$ [where here $f (x) = x^{2} / 2$]. 
This behaviour is in fact expected beyond the simple flat-potential example. 
The reason is the following. 
Let us consider the realisations that, starting from a certain initial field configuration $\vb*{\phi}$, realise an extremely small number of $e$-folds $\delta \mathcal{N}$. 
Over their course, the drift provides a contribution that schematically scales like $\delta \mathcal{N} \, \vb*{F}$, while the noise provides a contribution that scales like $\sqrt{ \delta \mathcal{N} } \, \vb*{G}^2$. 
Therefore, for sufficiently small $\delta\mathcal{N}$, the drift can be neglected, hence the lower tail of the first-passage-time distribution behaves as in free-diffusion problems, which feature L\'{e}vy distributions~\eqref{eq:Levy}. 
In \cite{Ezquiaga:2019ftu} for instance, the first-passage-time distribution was computed for linear and cubic potentials, and the same L\'{e}vy behaviour at small $\mathcal{N}$ was observed [with functions $f$ that depend on the model]. 
This behaviour is universal and guarantees that the attracting force becomes infinite when time approaches the target duration.\\

\noindent\textbf{\textit{Meanders}}~~
In meanders, the $Q$ function is given by the survival probability, which has already been computed in \Eq{eq:flat:survival}. 
This leads to 
\begin{equation}
     \widetilde{F} (x, \, N) 
     = \pdv{x} \ln \qty{ 
            \mathrm{erf} \qty[ 
                \frac{x}{\sqrt{2 (N_{{\mathrm{F}}} - N)}} 
            ] 
        } % \notag\\
     = \sqrt{ \frac{2}{ \pi (N_{{\mathrm{F}}} - N) }} \, \frac{
            \displaystyle 
            \exp \qty[ - \frac{x^2}{2 (N_{{\mathrm{F}}} - N)} ] 
     }{ 
            \displaystyle 
            \mathrm{erf} \qty[ \frac{x}{\sqrt{2 (N_{{\mathrm{F}}} - N)}} ] 
     } 
     \,\, . 
\end{equation}
When $N \leq N_{{\mathrm{F}}}$, close to the absorbing boundary one has $\widetilde{F} (x, \, N) = 1 / x + \mathcal{O} (x)$, hence the same repelling force as for excursions is at play. 
In the opposite limit, $x \gg N_{{\mathrm{F}}} - N$, the induced force decays, 
$\widetilde{F} (x, \, N) \simeq [ 2 / \pi (N_{{\mathrm{F}}} - N) ] \exp \qty[ -x^{2} / (N_{{\mathrm{F}}} - N) ]$, 
which contrasts with the excursion for which the induced force was attracting: this is because, in meanders, the condition $C$ does not impose to terminate inflation at $N_{{\mathrm{F}}}$.
 
When $N > N_{{\mathrm{F}}}$, in \Eq{eq:meandre:interm2} one has $P [ \mathcal{N} (\vb*{\phi}) \geq N_{{\mathrm{F}}} - N ] = 1$ since the duration of inflation is always positive. 
This leads the $Q$ function to equal $1$ in this regime, hence there is no induced force when $N > N_{{\mathrm{F}}}$ in stochastic meanders. 
Note that the induced force is continuous at the transition since from the above expression one can also check that $\widetilde{F} (x > 0, \, N \to N_{{\mathrm{F}}} - 0) = 0$.\\
 
\noindent\textbf{\textit{Bridges}}~~
In stochastic bridges, the absorbing boundary at $x = 0$ is removed [the field is free to fluctuate back and forth across the boundary], which leads to the simpler solution to the Fokker--Planck equation, 
\begin{equation}
    P (x, \, N \mid x_0, \, N = 0) 
    = \frac{1}{\sqrt{2 \pi N}} \, \exp \qty[ - \frac{ (x - x_{0})^2 }{2 N} ] 
    \,\, . 
    \label{eq:FP:sol:flat:noBoundary} 
\end{equation}
The $Q$ function is given by the backward Kolmogorov distribution, $Q(x, \, N) = P(x = 0, \, N \mid x, \, N = 0)$, and this leads to  
\begin{equation}
    \widetilde{F} 
    = - \pdv{x} \ln \qty{ 
        \frac{1}{2 \pi (N_{{\mathrm{F}}} - N)} \, \exp \qty[ - \frac{x^2}{2 (N_{{\mathrm{F}}} - N)} ] 
    } 
    = - \frac{x}{N_{{\mathrm{F}}} - N} 
    \,\, . 
\end{equation}
This corresponds to the attractive term in the excursion, see \Eq{eq:Finduced:excursion:bridge}, but there is no repelling term since the field is free to cross the boundary, $x = 0$, at any time.\\

Let us note that, in all three cases, the constrained distribution functions $\mathcal{P} (\vb*{\phi}, \, N \mid C)$ can be computed directly from \Eqs{eq:constrainedP:excursion}, \eqref{eq:constrainedP:meander} and \eqref{eq:constrainedP:bridge} respectively, and one can check that they satisfy the modified Fokker--Planck equations, with the expression of the induced drift derived in each case.
 
\section{S3.~Low-diffusion limit}
\setcounter{section}{3}
\setcounter{equation}{0}
 
In excursions and meanders, the induced force involves the first-passage-time distribution, which is most of the time not straightforward to derive since it implies solving the adjoint Fokker--Planck equation, which is a partial differential equation. 
In the limit where the amplitude of the stochastic noise in the unconstrained process~\eqref{eq:si_lan:SM} is small, the problem can nonetheless be simplified as follows. 
In the limit of low diffusion, the first-passage time distribution is approximately Gaussian, 
\begin{equation}
    P_{{\mathrm{FPT}}}  (\vb*{\phi}, \, \mathcal{N}) 
    \simeq \frac{1}{\sqrt{2 \pi \sigma^2 (\vb*{\phi})}} \, \exp \qty{ 
        - \frac{ [ \mathcal{N} - \mu (\vb*{\phi}) ]^2 }{ 2 \sigma^2 (\vb*{\phi}) } 
    } 
    \,\, , 
  \label{eq:Pfpt:Gaussian}
\end{equation}
where $\mu (\vb*{\phi}) \equiv \expval{ \mathcal{N} (\vb*{\phi})}$ is the mean number of $e$-folds realised from $\vb*{\phi}$ and $\sigma^2 (\vb*{\phi})\equiv \expval{ \mathcal{N}^2 (\vb*{\phi}) } - \expval{ \mathcal{N} (\vb*{\phi}) }^2$ is its variance. 
These moments can be obtained from noticing that the adjoint Fokker--Planck equation~\eqref{eq:fp_adj} yields~\cite{Vennin:2015hra, Vennin:2016wnk}
\begin{equation}
    \mathcal{L} \expval{ \mathcal{N}^{n} (\vb*{\phi}) } 
    = - n \expval{ \mathcal{N}^{n-1} (\vb*{\phi}) } 
    \,\, ,
\end{equation}
which can be solved iteratively. 
For the first two moments, it gives 
\begin{equation}
    \mathcal{L} \mu = -1
    \qquad \text{and} \qquad 
    \mathcal{L} \sigma^{2} = - \vb*{G}^{2} \vcentcolon ( \grad \mu ) \otimes ( \grad \mu ) 
    \,\, .
\end{equation}
At leading order in the amplitude of the noise, these equations reduce to $\vb*{F} \cdot \grad \mu = - 1$ and $\vb*{F} \cdot \grad \sigma^{2} = - \vb*{G}^2 \vcentcolon (\grad \mu) \otimes (\grad \mu)$. 
This implies that $\mu$ is of order $\vb*{G}^{0}$, $\sigma^2$ is of order $\vb*{G}^{2}$, and one can show that higher-order connected moments are of higher order in $\vb*{G}$. 
This explains why the first-passage-time distribution is approximately Gaussian in the low-diffusion limit. 
Let us now compute the induced force in the case of stochastic excursions and meanders [stochastic bridges are not considered since as argued above they are less relevant for inflation].\\

\noindent\textbf{\textit{Excursions}}~~
In stochastic excursions, one has
\begin{align}
    \Delta \vb*{F} 
    &= \vb*{G}^{2} \cdot \grad \ln P_{{\mathrm{FPT}}} (\vb*{\phi}, \, N_{{\mathrm{F}}} - N ) \notag \\ 
    &\simeq \frac{N_{{\mathrm{F}}} - N - \mu}{\sigma^2} \vb*{G}^{2} \cdot \grad \mu 
    + 
    \qty[ 
        \frac{(N_{{\mathrm{F}}} - N - \mu)^2}{2 \sigma^{4}} - \frac{1}{2 \sigma^{2}} 
    ] \vb*{G}^{2} \cdot \grad \sigma^2 
    \,\, .
\end{align}
The first two terms are of order $\sigma^{-2}$, while the third term is of order $\sigma^{0}$. 
The third term is thus suppressed by higher powers of the noise amplitude, and at the order at which the calculation is performed one has
\begin{equation}
    \Delta \vb*{F} 
    \simeq \frac{ (N_{{\mathrm{F}}} - N - \mu)^2}{2 \sigma^{4}}
    \vb*{G}^{2} \cdot \grad \sigma^{2} 
    + \frac{ N_{{\mathrm{F}}} - N - \mu }{\sigma^{2}} \vb*{G}^{2} \cdot \grad \mu 
    \,\, .
    \label{eq:induced:force:excursions:Gaussian}
\end{equation}
Recalling that $\sigma^{2}$ is of order $\vb*{G}^{2}$, it is worth noticing that the induced force is of order $\vb*{G}^{0}$, \ie it does not depend on the amplitude of the noise. 
At leading order in the amplitude of the noise, the constrained Langevin equation~\eqref{eq:si_lan:constrained:SM} thus reduces to
\begin{equation}
    \dv{\vb*{\phi}}{N} 
    = \vb*{F} (\vb*{\phi}) + \Delta \vb*{F} (\vb*{\phi}, \, N) 
    \,\, ,
    \label{eq:si_lan:constrained:Gaussian:SM}
\end{equation}
in which the stochastic noise has been dropped since it is of order $\vb*{G}^{2}$. 
One therefore reaches the important conclusion that in the low-diffusion limit, stochastic excursions follow deterministic trajectories, given by \Eq{eq:si_lan:constrained:Gaussian:SM}. 
From \Eq{eq:induced:force:excursions:Gaussian}, one also sees that the induced force vanishes when $\mu (\vb*{\phi}) = N_{{\mathrm{F}}} - N$, \ie on the unconstrained trajectories that lead to the required number of $e$-folds, which is consistent.\\

\noindent\textbf{\textit{Meanders}}~~
In stochastic meanders, one has to compute the survival probability
\begin{equation}
    S (\vb*{\phi}, \, N) 
    = \int_{N}^{\infty} \dd \mathcal{N} \, P_{{\mathrm{FPT}}} (\vb*{\phi}, \, \mathcal{N}) 
    \simeq \frac{1}{2} \mathrm{erfc} \qty[ 
        \frac{N - \mu (\vb*{\phi})}{\sqrt{2 \sigma^{2} (\vb*{\phi})}} 
    ] 
    \,\, .
\end{equation}
This gives rise to the induced force
\begin{align}
    \Delta \vb*{F} 
    &= \vb*{G}^{2} \cdot \grad \ln S (\vb*{\phi}, \, N_{{\mathrm{F}}} - N) \notag \\
    &= \sqrt{\frac{2 \sigma^{2}}{\pi (N_{{\mathrm{F}}} - N - \mu)^2}} \, \frac{ 
        \displaystyle 
        \exp \qty[ 
            - \frac{ (N_{{\mathrm{F}}} - N - \mu)^2}{ 2 \sigma^{2} } 
        ] }{ 
            \displaystyle 
            \mathrm{erfc} \qty( \frac{ N_{{\mathrm{F}}} - N -\mu }{ \sqrt{2 \sigma^{2}}} ) 
        } 
    \qty[ 
        \frac{ (N_{{\mathrm{F}}} - N - \mu)^2}{2 \sigma^{4}} 
        \vb*{G}^{2} \cdot \grad \sigma^{2} 
        + \frac{ N_{{\mathrm{F}}} - N - \mu }{\sigma^{2}} \vb*{G}^{2} \cdot \grad \mu 
    ] 
    \,\, .
\end{align}
One recognises, inside the square brackets, the induced force obtained for excursions, see \Eq{eq:induced:force:excursions:Gaussian}. 
The prefactor before the square brackets is a function of $z (\vb*{\phi}, N) \equiv (N_{{\mathrm{F}}} - N -\mu) / \sqrt{2 \sigma^{2}}$, and it is instructive to consider the following limits.

When $z \gg 1$, the prefactor approaches $1$, given that $\mathrm{erfc} (z) = \exp (- z^{2})[1 / \sqrt{\pi} \, z + \mathcal{O} (1 / z^{3})]$. 
As a consequence, when $z$ is large, the induced force is the same for both types of processes. 
This corresponds to the regime where the unconstrained mean trajectory originating from $\vb*{\phi}$ realises much less $e$-folds than the targeted duration. 
In that case, the effective force pushes the system towards the target trajectory at the same rate for excursions and for meanders. 

When $\abs{ z } \ll 1$, the prefactor scales as $1 / z$, but in that regime the induced force scales as $\sigma^{2}$, hence at the order at which the calculation is performed the induced force vanishes anyway. 

When $z \ll -1$, the prefactor is approximately $\exp(- z^{2}) / 2 \sqrt{\pi} \, z$, so it is highly suppressed. 
This implies that, for field configurations that realise much more $e$-folds than $N_{{\mathrm{F}}}$ in the unconstrained setup, the induced force is much weaker for meanders than for excursions. 
This makes sense since only a lower bound on the duration of inflation is imposed for meanders.

Let us stress that the same conclusions were reached in the example of the flat potential discussed above, namely the fact that the repelling component of the induced force is the same in excursions and in meanders, while the attracting component differs and is much stronger for excursions. 
The flat potential is an example that does not pertain to the low-diffusion regime: on the contrary, since the drift vanishes, the noise plays a dominant role. 
This shows that these findings are in fact very generic. 
It explains why, in the physical examples investigated in the main text, when imposing a large value for $N_{{\mathrm{F}}}$, similar results are obtained with excursions and meanders. 

Finally, let us discuss the regime of validity of the Gaussian approximation~\eqref{eq:Pfpt:Gaussian}. 
In \cite{Pattison:2017mbe, Ezquiaga:2019ftu, Ezquiaga:2022qpw}, it was shown that the upper tail of the first-passage-time distribution is generically exponential or even heavier, and thus strongly departs from a Gaussian profile. 
The transition between Gaussian and heavier profiles occurs around a scale $N_{\mathrm{c}} (\vb*{\phi})$ that is model dependent,
and which generically increases when decreasing the amplitude of the noise (\ie the Hubble parameter in the context of inflation). 
In the low-diffusion limit, it is therefore expected that the exponential tail plays a negligible role, but one should keep in mind that the above approximation is valid only in field-time regions where 
\begin{equation}
    \mu (\vb*{\phi}) \ll N_{\mathrm{c}} (\vb*{\phi}) - ( N_{{\mathrm{F}}} - N ) 
    \,\, .
\end{equation}
When this condition is not satisfied, large differences between excursions and meanders are also expected anyway. 

The Gaussian approximation also breaks down on the lower tail, \ie when $\mathcal{N} \ll \expval{ \mathcal{N} } (\vb*{\phi})$. 
This is necessarily the case since, as explained below \Eq{eq:effectivepot:excursion:flat}, $P_{{\mathrm{FPT}}} (\vb*{\phi}, \mathcal{N} = 0) = 0$, which is obviously violated by \Eq{eq:Pfpt:Gaussian}. 
This condition is however responsible for the divergence of the confining force when $N \to N_{{\mathrm{F}}}$ in the constrained process, see again the discussion below \Eq{eq:effectivepot:excursion:flat}. 
Therefore, in the Gaussian approximation, one may in principle sample realisations that terminate inflation before $N_{{\mathrm{F}}}$, which is forbidden. 
This is however not a problem for the following reason. 
In the case of excursions, the Gaussian approximation reduces to studying deterministic trajectories that are quickly attracted towards one of the unconstrained trajectories that produces the required number of $e$-folds. 
This guarantees that, at this order in the approximation, forbidden trajectories are not sampled. 
In the case of meanders, if $\mu (\vb*{\phi})$ is smaller than the target duration, it has been shown that the constrained process is equivalent to an excursion, and the argument above applies. 
If $\mu (\vb*{\phi})$ is larger than the target duration, then there is no danger to produce too short realisations either. 

One concludes that, for the purpose of the analysis carried out in the present work, the Gaussian approximation is a reliable tool in the low-diffusion limit. 

\section{S4.~Quadratic single-field inflation}
\setcounter{section}{4}
\setcounter{equation}{0}

In this section, the quadratic single-field model is considered, 
\begin{equation}
    V (\phi) = 12 \pi^2 \Mp^2 v_{0} \phi^2 
    \,\, , 
    \label{eq:quadratic:inflation:pot} 
\end{equation}
where the potential is written in terms of the parameter $v_{0}$, such that it matches the convention adopted in the main text [namely \Eq{eq:quadratic:inflation:pot} matches \Eq{eq:df_pot} when $\phi_{2}$ is set to $0$ and $\phi_{1}$ is replaced with $\phi$]. 
In the slow-roll regime, the characteristic function associated with the first-passage-time distribution of the unconstrained dynamics~\eqref{eq:si_lan:SM} can be obtained analytically, as shown in \cite{Pattison:2017mbe}, 
\begin{equation}
    \chi (\phi, \, t) 
    = \qty( \frac{x}{x_{{\mathrm{F}}}} )^{(1 - \alpha) / 2} \, 
    \frac{
        \displaystyle 
        {}_1 F_{1} \qty( 
            \frac{ \alpha - 1 }{4}, \, 1 + \frac{\alpha}{2}, \, - \frac{2}{v_{0} x^{2}} 
        ) }{ 
        \displaystyle 
        {}_1 F_{1} \qty( 
            \frac{\alpha - 1}{4}, \, 1 + \frac{\alpha}{2}, \, - \frac{2}{v_{0} x_{{\mathrm{F}}}^{2}} 
        ) 
        } 
        \,\, , 
    \label{eq:Pfpt:quadratic:2} 
\end{equation}
from which one has 
\begin{equation}
    P_{{\mathrm{FPT}}} (\phi, \, \mathcal{N}) 
    = \int_{-\infty}^{\infty} \frac{\dd t}{2 \pi} \, \chi(\phi, \, t) \, e^{-i t \mathcal{N}} \,\, , 
    \label{eq:Pfpt:quadratic:1} 
\end{equation} 
where $\alpha=\sqrt{1 - 8 i t / v_{0}}$, $x = \phi / \Mp$, and $x_{{\mathrm{F}}} = \sqrt{2}$ is the value of $x$ at the end of inflation, and ${}_1 F_{1}$ is the Kummer confluent hypergeometric function. 

\subsection{S4.1~Low-diffusion regime}
\begin{figure}
  \centering
  \includegraphics[width = 0.5\linewidth]{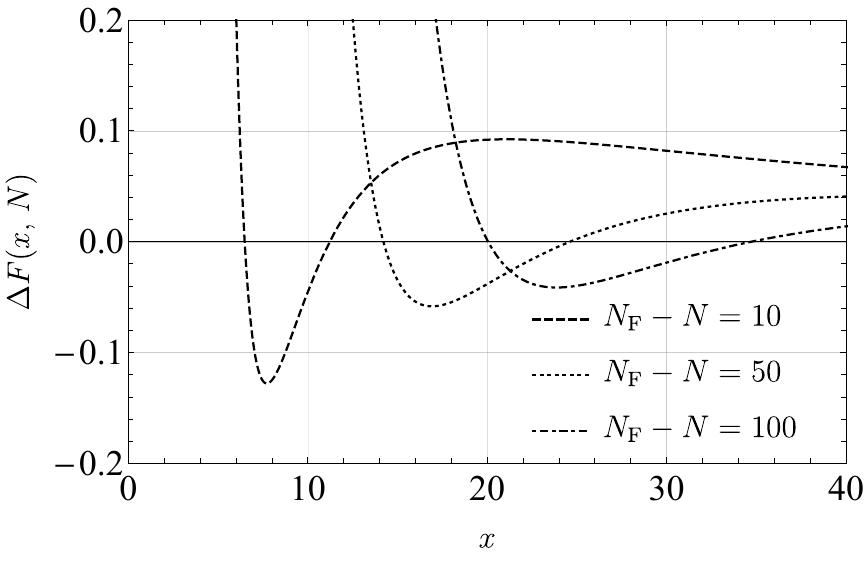}
  \caption{Induced force~\eqref{eq:indF_lowdifflim} in the quadratic single-field model, in the low-diffusion limit, at a few different times. When $\mu (x) > N_{{\mathrm{F}}} - N + (x^{6} - 8) / 6 x^{4}$, $\Delta F $ is positive (repelling induced force), which signals the breakdown of the Gaussian approximation. This has to be contrasted with the right panel of \Fig{fig:numerical:quadratic:singlefield} where this approximation is not performed.
  } 
  \label{fig:indF_Gaussian}
\end{figure}
When $v_0\ll 1$, \eqref{eq:Pfpt:quadratic:1}-\eqref{eq:Pfpt:quadratic:2} asymptote the Gaussian profile~\eqref{eq:Pfpt:Gaussian}, with $\mu (x) = (x^{2} - x_{{\mathrm{F}}}^{2}) / 4$ and $\sigma^{2} (x) = v_0 (x^{6} - x_{{\mathrm{F}}}^{6}) / 48$. 
In this limit the constrained dynamics becomes deterministic and the induced force~\eqref{eq:induced:force:excursions:Gaussian} acting on $x$ reads
\begin{equation}
    \Delta F (x, \, N) 
    = [ N_{{\mathrm{F}}} - N - \mu (x) ]^2 \frac{144 x^7}{(x^6 - 8 )^2} + [ N_{{\mathrm{F}}} - N - \mu (x) ] \frac{24 x^3}{x^6 - 8} 
    \,\, , 
    \label{eq:indF_lowdifflim}
\end{equation}
which is displayed in Fig.~\ref{fig:indF_Gaussian}. 
When $x$ is such that $\mu(x) < N_{{\mathrm{F}}} - N$, \ie when $x < \sqrt{2 + 4 (N_{{\mathrm{F}}} - N)}$, one has $\Delta F > 0$. 
This is consistent with the fact that, when the classical number of $e$-folds realised from $x$ is smaller than the target $N_{{\mathrm{F}}} - N$, the constrained dynamics selects realisations of the noise that kick the inflaton upwards in its potential, hence that increase the duration of inflation. 
When $x$ is such that $N_{{\mathrm{F}}} - N < \mu (x) <  N_{{\mathrm{F}}} - N + (x^6 - 8) / 6 x^4$, $\Delta F < 0$, which corresponds to the fact that realisations that would produce too many $e$-folds in the unconstrained setup are pushed towards smaller-field values in the constrained theory. 
However, when $x$ is such that $\mu (x) > N_{{\mathrm{F}}} - N + (x^6 - 8) / 6 x^4$, \ie for the largest values of $x$, the effective force becomes positive again, which clearly signals the breakdown of the Gaussian approximation. 
Indeed, when $\mu (x)$ becomes large, the first-passage-time distribution has to be evaluated in its lower tail when computing the induced force, and as explained above the Gaussian approximation necessarily fails in its lower tail since it predicts that $P_{{\mathrm{FPT}}}(x, \, \mathcal{N} = 0) > 0$ instead of the expected L\'{e}vy-like behaviour. 
In passing, this implies that even in the limit $v_{0} \to 0$, the regime of validity of the classical limit is limited, which is not obvious a priori~\cite{Vennin:2016wnk, Pattison:2017mbe}. 

As long as one remains far from this problematic regime, the Gaussian approximation can nonetheless be used, and in \Fig{fig:sf_traj} it gives rise to the constrained realisations that all converge to the unconstrained one well before the scales accessed in cosmological measurements are produced. This guarantees that selection effects do not affect the observable predictions of the model, as long as $N_{\mathrm{F}}$ is sufficiently large.  
\subsection{S4.2~Exact treatment}
\begin{figure}
  \centering
  \includegraphics[width = 0.95\linewidth]{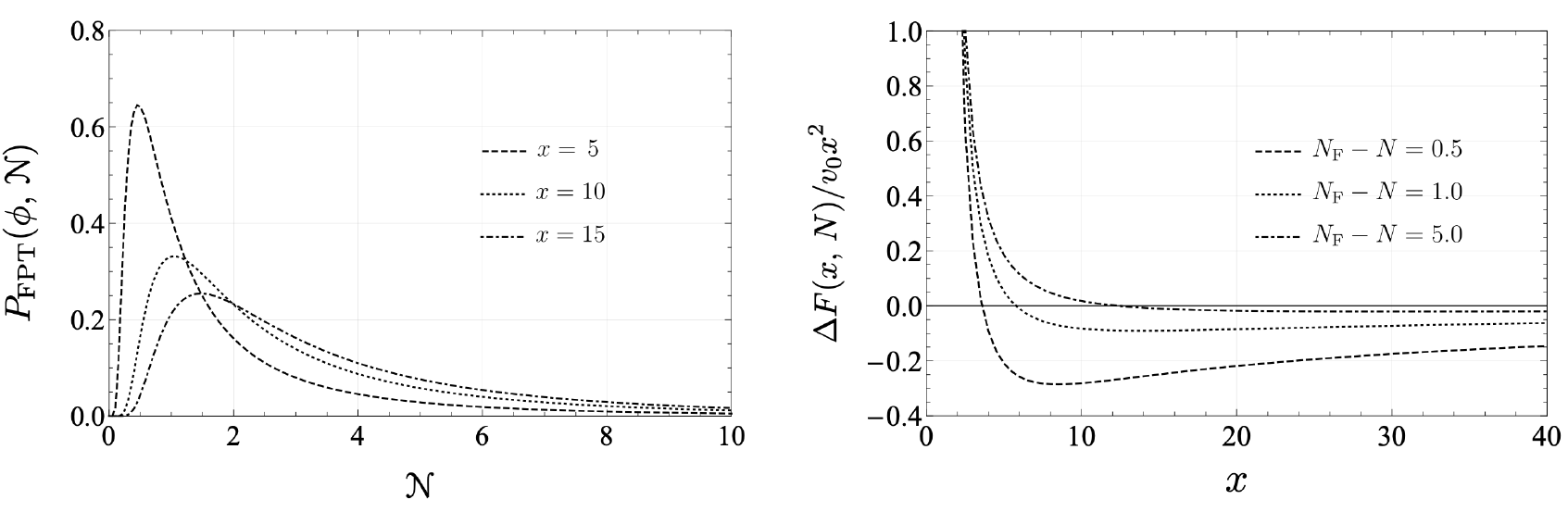}
  \caption{
    The full behaviour of $P_{{\mathrm{FPT}}} (\phi, \, \mathcal{N})$ (\textit{left}) and the induced force (\textit{right}), for the quadratic inflationary model (\ref{eq:quadratic:inflation:pot}) with $v_{0} = 1$. 
    Contrary to Fig.~\ref{fig:indF_Gaussian}, one can see that $\Delta F$ remains negative at large $x$.  
  } 
  \label{fig:numerical:quadratic:singlefield}
\end{figure}
When $v_{0}$ is not too small, or in the regime where $\mu (x) >  N_{{\mathrm{F}}} - N + (x^{6} - 8) / 6 x^{4}$, the Gaussian approximation is not sufficient and one must use the full result~\eqref{eq:Pfpt:quadratic:2}-\eqref{eq:Pfpt:quadratic:1}. 
The first-passage-time distribution is displayed in \Fig{fig:numerical:quadratic:singlefield}, together with the induced force. In practice, the latter is computed using that
\begin{equation}
    \Mp \pdv{P_{{\mathrm{FPT}}} (\phi, \, \mathcal{N})}{\phi} 
    = 
    \frac{1}{x}  
    \int_{- \infty}^{\infty} \frac{\dd t }{2 \pi} \, e^{- i t \mathcal{N}} \, 
    \qty( \frac{x}{x_{{\mathrm{F}}}} )^{(1 - \alpha) / 2} \, 
    \frac{
        \displaystyle 
        \frac{1 + \alpha}{2} \, {}_1 F_{1} \qty( \frac{\alpha - 1}{4}, \, 1 + \frac{\alpha}{2}, \, - \frac{2}{v_{0} x^2} ) - \alpha \, {}_1 F_{1} \qty( \frac{\alpha - 1}{4}, \, \frac{\alpha}{2}, \, - \frac{2}{v_{0} x^2} ) 
    }{
        \displaystyle 
        {}_1 F_{1} \qty( \frac{\alpha - 1}{4}, \, 1 + \frac{\alpha}{2}, \, - \frac{2}{v_{0} x_{{\mathrm{F}}}^2} ) 
    } \,\, , 
\end{equation}
where the identity $\displaystyle z \frac{\partial}{\partial z}\ {}_1 F_{1} (a, \, b, \, z) = (b - 1) [ {}_1 F_{1} (a, \, b-1, \, z) - {}_1 F_{1} (a, \, b, \, z) ]$ has been employed. 
One can check that, at large $x$, the induced force remains negative, which confirms that the problem discussed above is an artefact of the Gaussian approximation. 

\begin{figure}
  \centering
  \includegraphics[width = 0.5\linewidth]{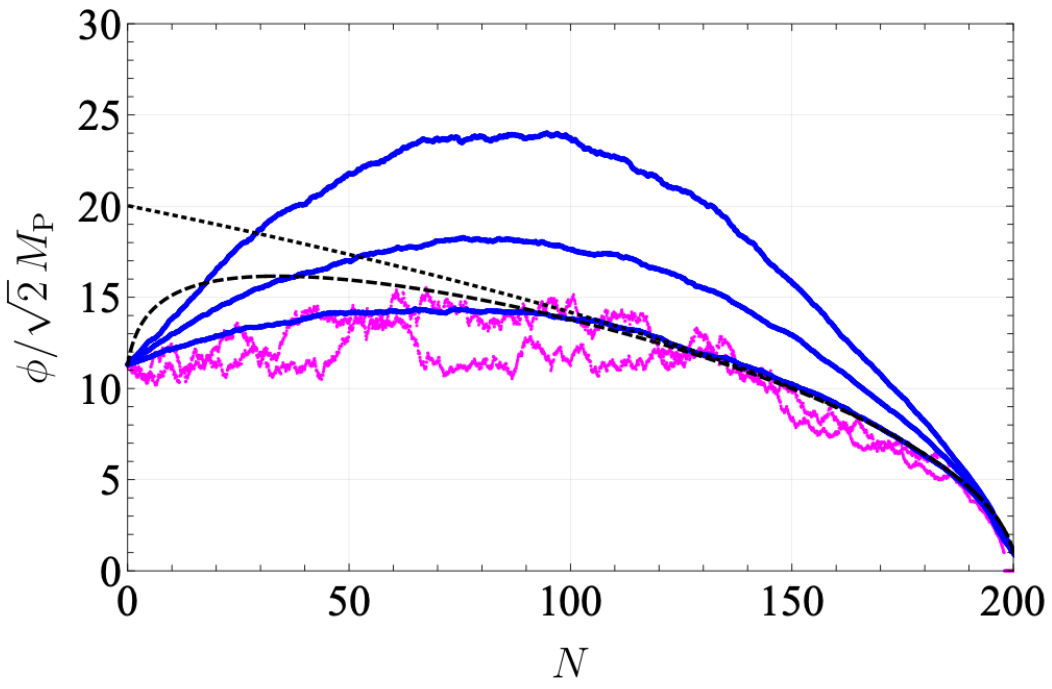}
  \caption{Constrained dynamics in the quadratic single-field model~\eqref{eq:quadratic:inflation:pot}. The blue lines stand for the mean constrained trajectories, starting from $\phi_{0} / \Mp = 16$ and realising $N_{{\mathrm{F}}} = 200$, for three different values of $v_{0}$ ($v_{0} = 10^{-3}$, $5 \times 10^{-3}$, and $10^{-2}$, from bottom to top). In practice, $2 \times 10^{5}$ unconstrained realisations are produced, and the average trajectory is computed over the ones that record $N_{{\mathrm{F}}} \in [195, \, 205]$ (namely 127 realisations when $v_{0} = 10^{-3}$, 725  when $v_{0} = 5 \times 10^{-3}$, and 859 when $v_{0} = 10^{-2}$). Two examples of individual realisations are displayed in magenta in the case $v_{0} = 10^{-3}$. The black dashed line corresponds to the constrained realisation that produces $N_{{\mathrm{F}}} = 200$ in the Gaussian approximation, which applies in the limit $v_0\ll 1$ and yields a deterministic dynamics. Finally, the black dotted line stands for the unconstrained, classical (\ie in the absence of noise) trajectory that terminates at $N_{{\mathrm{F}}} = 200$.}
  \label{fig:traj:quadratic:singlefield}
\end{figure}

In \Fig{fig:traj:quadratic:singlefield}, constrained realisations with $N_{{\mathrm{F}}} = 200$ are displayed from the initial condition $x = 16$ and for a few values of $v_{0}$. 
These values are much larger than what is allowed by the measurement of the temperature anisotropies in the CMB, which rather impose $v_{0} < 10^{-13}$, but they have been used in order to study the behaviour of the constrained dynamics in cases where quantum diffusion plays a more prominent role in the unconstrained setup. 
One can see that, compared to \Fig{fig:sf_traj}, the profile of the detour is less sharp: instead of occurring mostly at early time, it now unfolds during the whole inflationary stage. 
This is consistent with the results obtained in \cite{Gratton:2005bi, Gratton:2010zv}. When $v_{0}$ decreases, the maximum deviation from the unconstrained trajectory takes place at earlier time and is of smaller amplitude. 
This confirms that, in the low-diffusion regime $v_{0} \ll 1$, most of the selection effects take place at early time since this is when the amplitude of the noise is the largest, hence this is when the occurrence of large stochastic fluctuations that divert the system away from the unconstrained mean path is the most likely. 

\section{S5.~Distribution of the field ratio in quadratic double-field inflation}
\setcounter{section}{5}
\setcounter{equation}{0}

In the main text, it is shown that the stochastic selection effect diverts the field-space trajectories towards single-field attractors. 
The size of the corresponding blue region in the constrained setups (the middle and right panels in \Fig{fig:df_traj}) is increased compared to the unconstrained ones (left panels in the same figure). 
One may wish to further quantify this effect, and determine how likely single-field phenomenology emerges at CMB scales from a randomly chosen initial condition. 

This section thus aims at providing the probability distribution of the indicator $\lambda$, observing the selection effect from a point of view that is complementary to \Fig{fig:df_traj}. In practice, $\lambda$ is a function of the field ratio $w = y/x$, see \Eq{eq:lambda:w}, so \Fig{fig:df_hist} shows the distribution of $w$ evaluated $60$ $e$-folds before the end of inflation. 
The mass ratio is fixed to $\sqrt{r} = m_{2} / m_{1} = 2$. 
The initial conditions are randomly chosen on the $\expval{ \mathcal{N} } = 60$ contour, and the equations of motion are solved for each initial condition and for each value of $N_{\mathrm{F}}$, to give rise to the distribution. 
The black line corresponds to the classical trajectory that realises $\expval{ \mathcal{N} } = 60$, while the other lines are obtained from the constrained trajectories with several $N_{\mathrm{F}}$'s.

One can see that, in the unconstrained setup, the distribution peaks at $\ln w \sim 0$, or equivalently $x \sim y$, hence multiple-field effects are likely to be observed. However, this simply reflects our choice of a flat prior on the angular coordinate of the initial field values and different results would be obtained with different priors. It nonetheless provides a benchmark distribution to which constrained setups can be compared. Indeed, when selection effects are implemented, a second peak appears at smaller $w$. When $N_{\mathrm{F}}$ increases, this peak shifts to lower values of $w$, while the local maximum around $\ln w\sim 0$ gets more and more suppressed. The appearance of this secondary peak is not caused by our choice of prior, which rather disfavours low values of $w$. Therefore we conclude that selection effects make low values of $w$, hence single-field phenomenology, more likely. This is consistent with the discussion surrounding \Fig{fig:df_traj}.

\begingroup
\begin{figure}[!h]
  \centering
  \includegraphics[width = 0.85\linewidth]{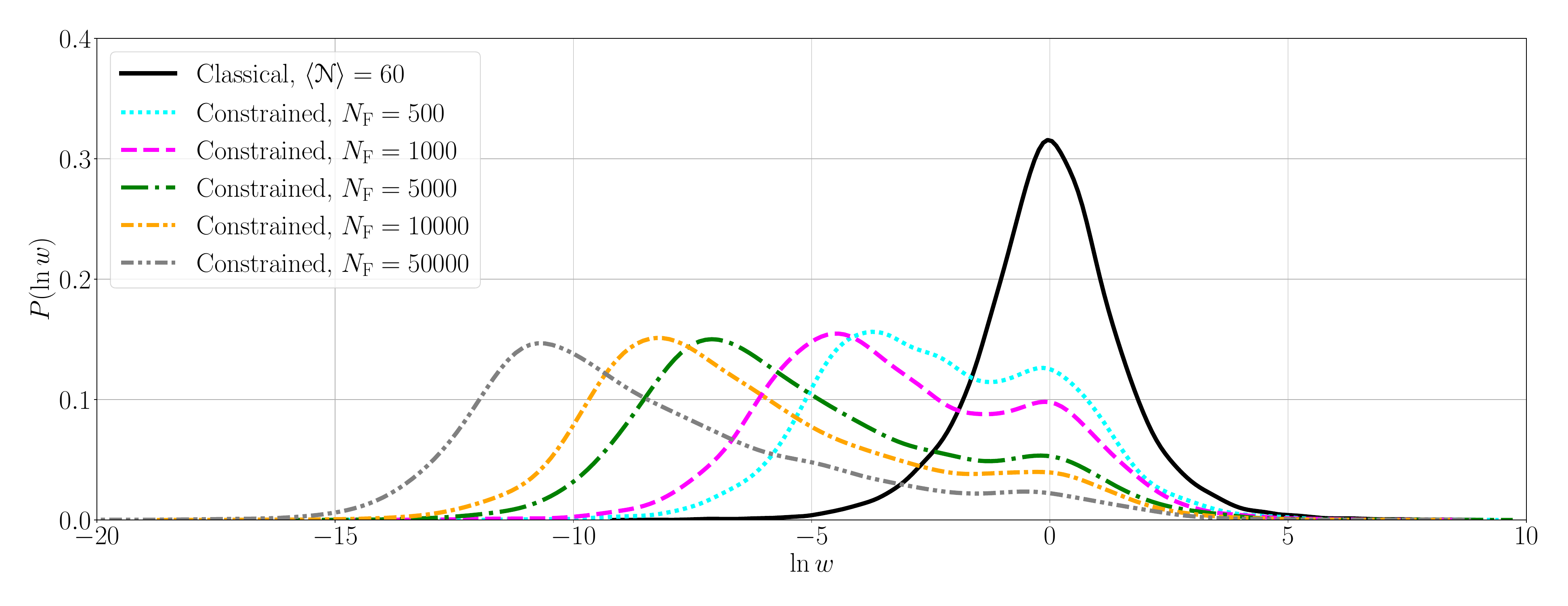}
  \caption{
  	Probability distribution of $w=\phi_2/\phi_1$, $60$ e-folds before the end of inflation, in double quadratic inflation with $m_2=2m_1$. In practice, $10^4$ initial conditions are drawn along the contour $\langle \mathcal{N} \rangle =60$ (with a flat prior on the field angular coordinate). From each initial condition, the constrained dynamics is generated and the value of $w$ is recorded $60$ e-folds before the end of inflation. The distribution of $w$ is reconstructed from this set of values, and different curves correspond to different values of $N_{\mathrm{F}}$.
  }
  \label{fig:df_hist}
\end{figure}

\end{document}